\documentclass[fleqn,10pt]{wlscirep} \usepackage[utf8]{inputenc} \usepackage[T1]{fontenc}
\usepackage{lineno} 

\title{A high sensitivity Cherenkov detector for Prompt Gamma Timing and Time Imaging}
\author[1]{Maxime Jacquet} \author[1]{Saba Ansari} \author[1]{Marie-Laure Gallin-Martel}
\author[1]{Ad\'{e}lie Andr\'{e}} \author[2]{Yannick Boursier} \author[2]{Mathieu Dupont}
\author[3]{Jilali Es-smimih} \author[1]{Laurent Gallin-Martel} \author[4]{Jo\"{e}l H\'{e}rault}
\author[1]{Christophe Hoarau} \author[4]{Johan-Petter Hofverberg} \author[4]{Daniel Maneval}
\author[2]{Christian Morel} \author[1]{Jean-Fran\c{c}ois Muraz} \author[3]{Fabrice Salicis}
\author[1,*]{Sara Marcatili}
\affil[1]{ Universit\'{e} Grenoble Alpes, CNRS, Grenoble INP, LPSC-IN2P3 UMR 5821, 38000 Grenoble,
France} \affil[2]{Aix-Marseille Univ, CNRS/IN2P3, CPPM, Marseille, France} \affil[3]{Ion beam
application SA, 3, chemin du Cyclotron, 1348 Louvain-La-Neuve, Belgium} \affil[4]{Centre Antoine
Lacassagne, 06200 Nice, France}
\affil[*]{sara.marcatili@lpsc.in2p3.fr}

\begin{abstract} 
We recently proposed a new approach for the real-time monitoring of particle therapy treatments with the goal of achieving high sensitivities on the particle range measurement already at limited counting statistics. This method extends the Prompt Gamma (PG) timing technique to obtain the PG vertex distribution from the exclusive measurement of particle Time-Of-Flight (TOF). It was previously shown, through Monte Carlo simulation, that an original data reconstruction algorithm (Prompt Gamma Time Imaging) allows to combine the response of multiple detectors placed around the
target. The  sensitivity of this technique depends on both the system time resolution and the beam intensity. 
At reduced intensities (Single Proton Regime - SPR), a millimetric proton range sensitivity can be achieved, provided the overall PG plus proton TOF can be measured with a 235 ps (FWHM) time resolution. 
At nominal beam intensities, a sensitivity of a few mm can still be obtained by increasing the number of incident protons included in the monitoring procedure. \\ 
In this work we focus on the experimental feasibility of PGTI in SPR through the development of a multi-channel, Cherenkov-based PG detector with  a
targeted  time resolution of 235 ps (FWHM): the TOF Imaging ARrAy (TIARA). Since PG emission is a rare phenomenon, TIARA  design is led by the concomitant optimisation of its detection efficiency and Signal to Noise Ratio (SNR). The PG module that we developed is composed of a small  PbF$_{2}$ crystal coupled to a silicon photoMultiplier  to provide the time stamp of the PG. This module is currently read in time coincidence with a diamond-based beam monitor placed upstream the target/patient to measure the proton time of arrival. TIARA will be eventually composed of 30 identical modules uniformly arranged around the target. The absence of a collimation system and the use of Cherenkov radiators are both crucial to  increase the detection efficiency and the SNR, respectively.\\
 A first prototype of the TIARA block detector was tested with 63 MeV protons delivered from a cyclotron: a  time resolution of 276 ps (FWHM) was obtained, resulting in a proton range sensitivity of 4 mm at 2$\sigma$ with the acquisition of only 600 PGs. A second prototype was also evaluated with 148 MeV protons delivered from a synchro-cyclotron obtaining a time resolution below 167 ps (FWHM) for the gamma detector. Moreover, using two identical PG modules, it was  shown that a uniform sensitivity on the PG profiles would be achievable  by combining the response of gamma detectors uniformly distributed around the target.\\ This work provides the experimental proof-of-concept for the development of a high sensitivity detector that can be used to  monitor particle therapy treatments and potentially act in real-time if the irradiation does not comply to
treatment plan.
\end{abstract}
\begin{document} \flushbottom \maketitle \thispagestyle{empty}
\section*{Introduction} Protons have a very peculiar dose deposition profile compared to X-rays,
with a sharp maximum at the end of their range (called Bragg peak), a limited entrance dose and
nearly zero dose deposition after the Bragg peak \cite{durante_nuclear_2016}. These figures seem
ideal to deliver highly conformal irradiations with a reduced number of fields and the highest
selectivity. However, while the physical advantage of Proton Therapy (PT) is widely acknowledged, PT
is still far from reaching its full potential. One of the most compelling open issues in PT is the
reduction of irradiation uncertainties \cite{pausch_detection_2020, parodi_vivo_2018}. The main
reason for this holdup comes from the objective technical complexity of predicting and verifying the
proton path in the patient.\\
Some imaging devices have been proposed in the past to monitor the proton range \textit{in vivo},
which are modified versions of classical nuclear medicine imaging modalities
\cite{park_multi-slit_2019, priegnitz_detection_2016,  llosa_first_2013, thirolf_development_2014,
parodi_vision_2015, bisogni_inside_2016, ferrero_innovation_2019}, while other more original
approaches do not provide  the full PG profile  (e.g.  prompt gamma
spectroscopy\cite{hueso-gonzalez_full-scale_2018}, PG integral
counts\cite{krimmer_cost-effective_2017, hueso-gonzalez_compact_2020}). The existing methods have
been extensively described in multiple reviews \cite{krimmer_prompt-gamma_2018, parodi_vivo_2018,
kraan_range_2015}. All these approaches exploit the spatial and/or temporal and/or energetic correlation 
of secondary particles emitted as a result of proton interactions with the biological tissue
\cite{min_prompt_2006, parodi_vivo_2018}. Among them, those based on the detection of Prompt Gamma
(PG) rays resulting from non-elastic nuclear collisions in the patient are of particular interest to
achieve a real-time measurement of the proton range and to allow stopping/correcting the treatment
procedure from its very beginning in case a deviation from treatment planning is detected.  Thanks to
the prompt nature of these emissions, it is in principle possible to accomplish a statistically
significant measurement of the proton range from the  first irradiation spot.
However,  PG emission being a rare phenomenon (PG yield is $\lesssim$10$^{-2}$~PGs/proton/cm, with
significant variations among models\cite{wronska_prompt-gamma_2021}), the development of high
sensitivity PG detectors is crucial to obtain a real-time information. The required sensitivity can
only be achieved by concurrently improving the intrinsic resolution of the measurement at a single
event scale, and  increasing the system detection efficiency to boost the measurement statistics. \\
With these goals in mind we have recently proposed a new modality to measure the PG vertex
distribution \textit{in vivo}: Prompt Gamma Time Imaging (PGTI)
\cite{jacquet_time--flight-based_2021}. With PGTI,  the overall Time-Of-Flight (TOF) of the proton
($T_{p}$) followed by the PG ($T_{PG}$) is first measured, as in the conventional PG Timing (PGT)
approach \cite{golnik_range_2014, hueso-gonzalez_first_2015}. Then, as both $T_{p}$ and $T_{PG}$
depend on the PG vertex coordinates ${\bf r}_v$, the PGTI reconstruction algorithm retrieves the
latter from the following equation: 
\begin{linenomath*} 
\begin{equation} TOF=T_{p}({\bf r}_v)+T_{PG}({\bf r}_v,{\bf
r}_d) \label{eq:pgti} 
\end{equation} 
\end{linenomath*}
where $TOF$ is experimentally measured and ${\bf r}_d$ are the
PG hit coordinates at the detector level, or the detector position coordinates if the detector is
small enough. Briefly, PGTI allows to convert the measured TOF distribution into a spatial
distribution, by performing an event-by-event deconvolution of the PG TOF $T_{PG}({\bf r}_v,{\bf
r}_d)$. The direct consequence of this approach is to enable the readout of multiple detectors
evenly distributed around the target: a capability that can be exploited to build a monitoring
system that is sensitive to proton beam deviations both along the beam axis and in the transverse
plane. The expected sensitivity of this technique has already been evaluated through Monte Carlo
(MC)\cite{jacquet_time--flight-based_2021} for different beam intensities. \\
 A PGT-based detector 
naturally  offers a high detection efficiency (of the order of $10^{-3}$) as no collimation system
is needed, and it is therefore a good candidate for real-time monitoring. In order to push the
system TOF resolution, and therefore its spatial resolution, we have proposed  to use a fast,
diamond-based beam monitor operated in single proton regime \cite{dauvergne_role_2020} to tag in
time each incident proton separately. Such a system, read-out in time coincidence with a
conventional gamma detector, has already allowed us to reach TOF resolutions of 237 ps (FWHM) in a
previous experiment with 68 MeV protons \cite{marcatili_ultra-fast_2020}.\\ Our current focus is  to
develop a fast gamma detector dedicated to PGTI\cite{marcatili_100_2019}. The TIARA (TOF Imaging
ARrAy) detector will be composed of approximately 30 independent modules evenly distributed around
the target to achieve 4$\pi$ coverage. Each module will be composed of a $\sim$1 cm$^3$ monolithic
Cherenkov radiator (PbF$_2$) read by Silicon PhotoMultipliers (SiPM). This system is designed to
allow the measurement of the PG time of arrival with excellent time resolution, and its hit position
with a  spatial resolution limited to  the Cherenkov crystal size. The use of a pure Cherenkov
radiator offers several advantages compared to the use of more conventional scintillation detectors.
The Cherenkov emission process is inherently faster than the scintillation, which  favours  temporal
resolution. At the same time, Cherenkov radiators generally have higher effective Z than
scintillators, which improves photon interaction probability. Nevertheless, we will show that their
greatest advantage lies in their relatively insensitivity to neutrons, which allows to set a natural
cut-off on one of the largest sources of background in proton-therapy monitoring, therefore increasing the 
Signal to Noise Ratio (SNR) of the measurement and reducing signal pile-up.\\ We have shown by MC
simulation how this detector can be operated in three different regimes depending on the beam intensity\cite{jacquet_time--flight-based_2021}. 
In Single Proton Regime (SPR), the beam intensity is reduced during
the first irradiation spot to approximately 1 proton per bunch. Under this regime, the statistics 
available for the proton range
measurement is limited, but the excellent time resolution achievable (of the order of 235 ps FWHM)
ensures a measurement sensitivity of the order of 1 mm at 2$\sigma$ for 10$^{8}$ protons of 100 MeV
and a simulated detection efficiency of 0.6\%. Alternatively, a sensitivity of 3 mm at 2$\sigma$ was
also predicted for 10$^{7}$ incident protons. \\ 
At nominal beam intensity (from $\sim$2000 to $\sim$2 million protons every 16 ns at isocenter\cite{henrotin_commissioning_2016} for the clinical accelerator used in this work), the time resolution is
ultimately limited by the bunch time-width of the proton beam as it is impossible, with the current
beam monitor, to establish which proton in the bunch has generated the  detected PG. Nevertheless,
the loss of time resolution is substantially compensated by the increased measurement statistics,
and sensitivities of a few mm can still be reached, depending on the number of PGs included in the
monitoring procedure: in our previous, work we estimated a proton range sensitivity of 2 mm (at 2$\sigma$)
for 10$^{9}$ incident protons\cite{jacquet_time--flight-based_2021}. \\ Finally, at very high intensities, a
one-value measurement of the proton beam displacement can still be obtained by computing the centre
of gravity of the TIARA detection modules. With this approach, no assumption should be made on the
detector time resolution as only the counts registered in each module (and their absolute position)
are needed in the reconstruction formula. A sensitivity of 2 mm (at 2$\sigma$) on the proton beam lateral
 displacement was estimated for 10$^{8}$ incident protons, whereas the method is less sensitive to distal proton range shifts.\\ 
 The present work completes our previous simulation
studies by confirming the hypotheses  made and demonstrating the experimental feasibility of PGTI in SPR. In this case, the reduced intensity regime allows to characterise the inherent performances of the detection module, 
 that would otherwise be affected by the time properties of the accelerator.
Through the results obtained in two experiments carried out with 63 and 148 MeV protons, the
performances of the detector in terms of time resolution will be presented in the first section. We
will also  show  that the gamma-ray energy measurement can be disregarded when using  extremely fast
detectors, and how a millimetric sensitivity on  proton range measurement could be achieved with an
unprecedentedly low PG statistics. In the second section, we will show the advantages of performing
PGTI with detectors placed at multiple angular positions with the aim to reach a uniform sensitivity
in the whole field of view. Finally, the different sources of background affecting our
Cherenkov-based detector will be discussed in the third section.
\section*{Results}
\subsection*{Detector characterisation with 63 MeV protons }
\begin{figure}[!ht] \centering \includegraphics[width=0.75\textwidth]{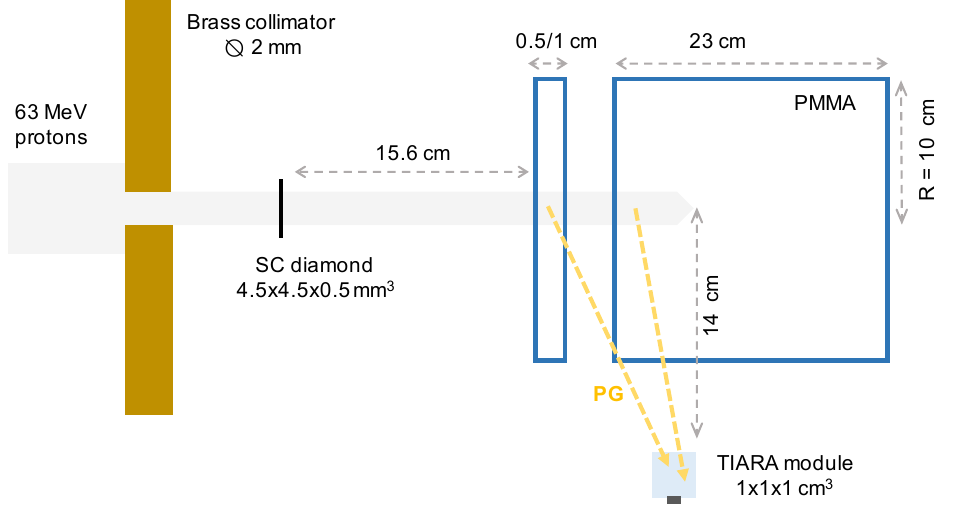}
\caption{Set-up of the experiment carried out at the MEDICYC facility with 63 MeV
protons. A first version of the TIARA module, composed of a 1 cm$^3$ PbF$_2$ crystal coupled to a
3$\times$3 mm$^2$ SiPM and facing the Bragg peak region at 90$^{\circ}$, was tested in time coincidence with a sc~diamond detector of 
$4.5\times4.5\times0.55$ mm$^3$ volume. For the time and
energy response characterisation, only PG signals from the thin PMMA target (cylinder of 10 cm radius) 
were considered: repeated measurement were carried out with either a 5 mm or a 1 cm thick target and 
the results were then averaged. The second target (cylinder of 10 cm radius and 23 cm thickness) was 
employed in the measurement of the proton range sensitivity;
initially placed at 3 cm distance from the thin target, it was progressively moved by 2, 4, 6 and 10
mm in order to induce an artificial shift in the proton range. All the targets have a density of 1.19 g/cm$^3$.
 } \label{fig:setup1_63} \end{figure}

\begin{figure}[!ht] \centering \includegraphics[width=0.49\textwidth]{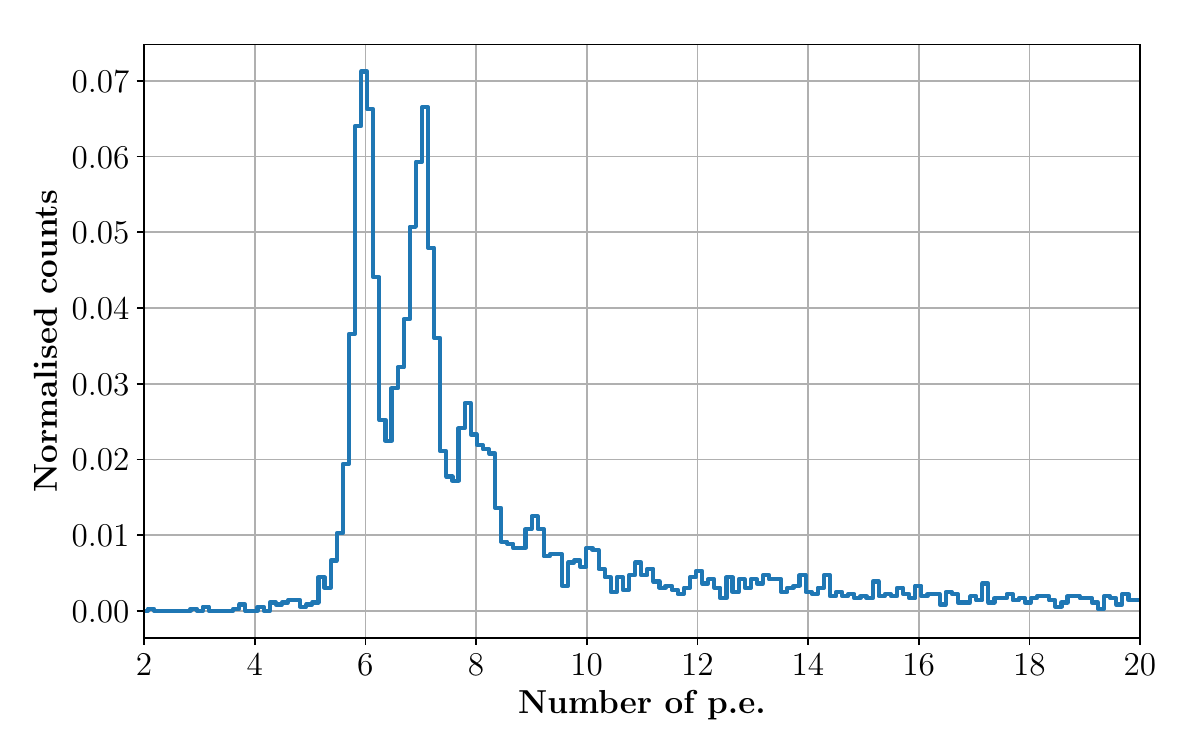}
\includegraphics[width=0.49\textwidth]{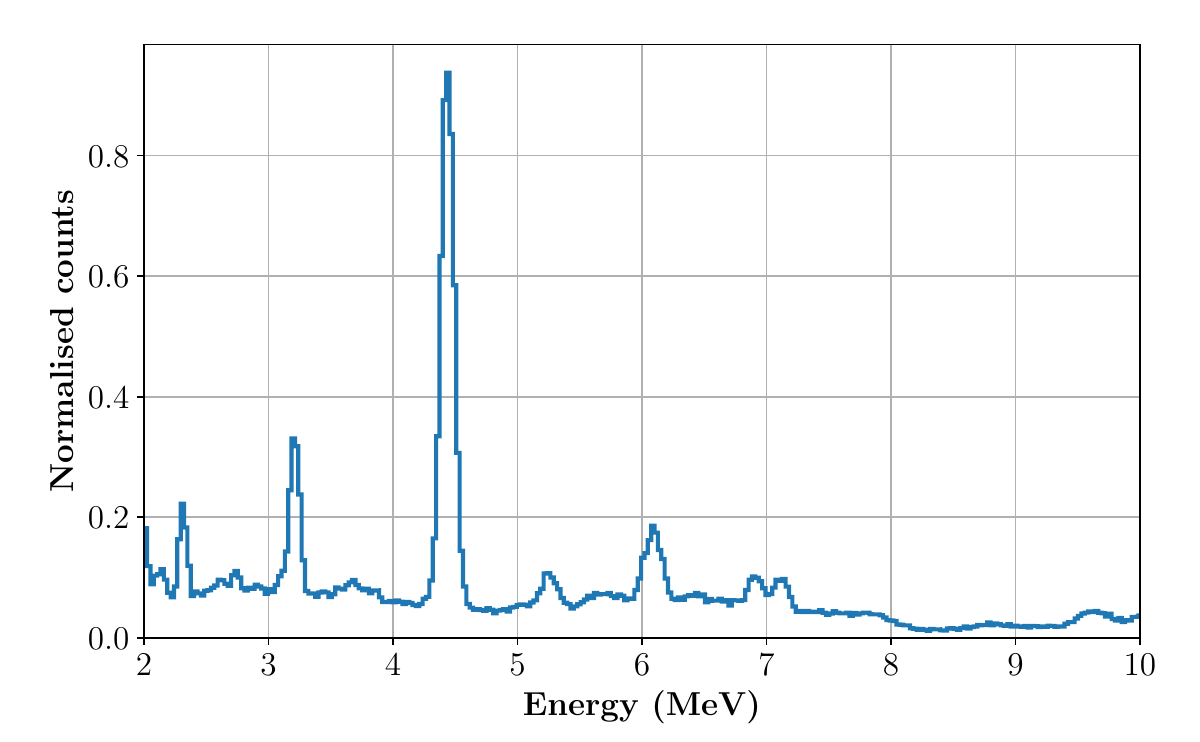} \caption{The left plot shows the histogram
of the energy deposited in the TIARA module (1 cm$^3$ PbF$_2$ crystal coupled to a
3$\times$3 mm$^2$ SiPM and placed at 90$^{\circ}$  
respect to the beam) expressed as the integral of the SiPM signal. PGs are generated by 63 MeV 
protons impinging on the two PMMA targets described in figure \ref{fig:setup1_63}. The first
peak in the histogram corresponds to 6 p.e; the 7, 8, 9 p.e. peaks are also visible: a median number of 7 p.e. per PG were detected. The right plot shows the expected energy spectrum of PGs obtained by MC 
simulation  for comparison. The simulation
is performed with the Geant4.10.4.p02 toolkit implementing the QGSP-BIC-EMY
physics list and reproducing the PMMA target and beam parameters used in the experiment. } \label{fig:energy}
\end{figure}

\begin{figure}[!ht] \centering 
\includegraphics[width=0.49\textwidth]{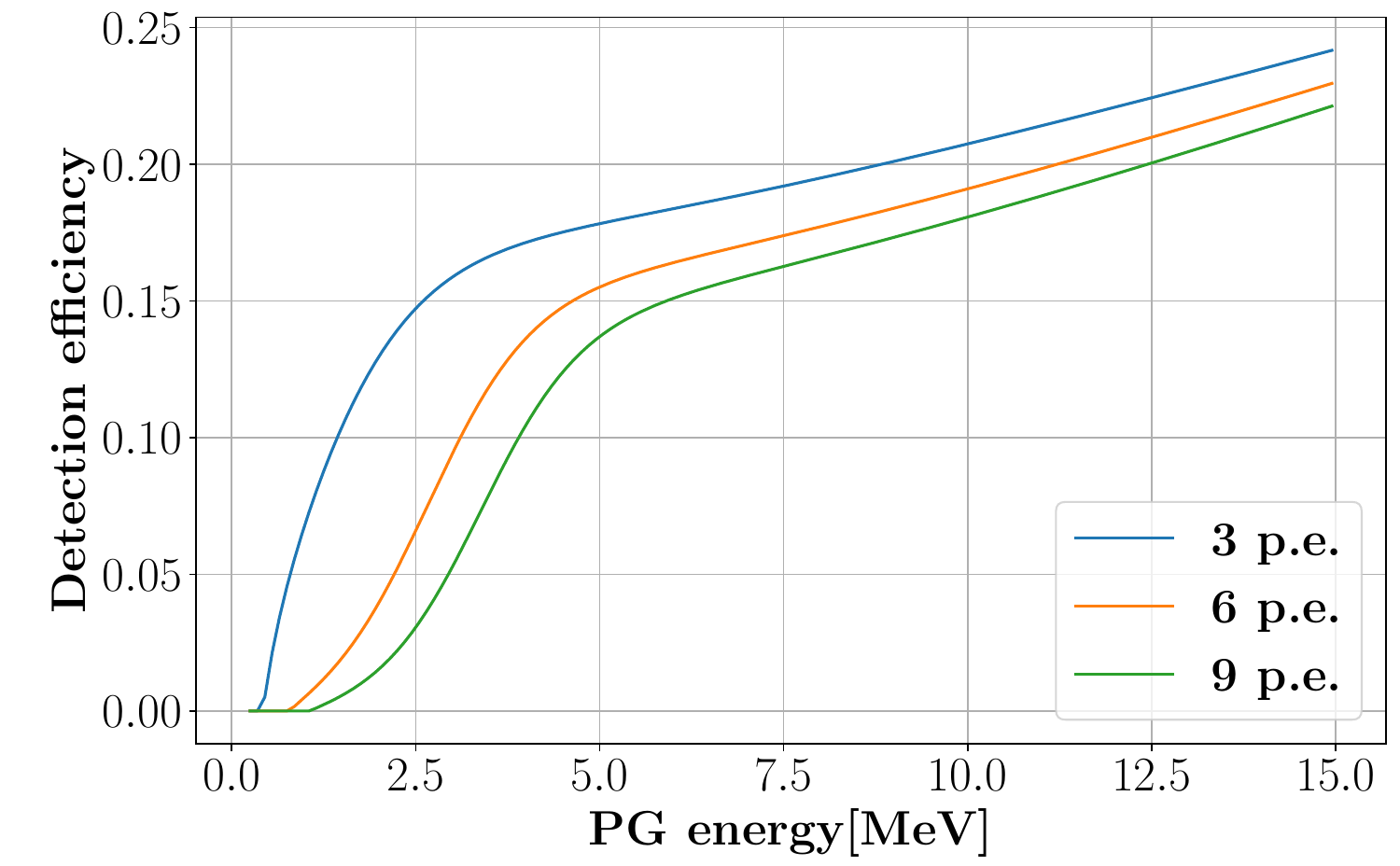} 
\caption{
Intrinsic detection efficiency (neglecting geometrical efficiency) of a PG 
detection module composed of a 1~cm$^3$ PbF$_2$ crystal coupled to a 3$\times$3 mm$^2$ SiPM.
The three functions were obtained from MC simulation using the Geant4.10.4.p02 toolkit with the QGSP-BIC-EMY
physics list  to establish the PG interaction probability, while the UNIFIED model 
was applied to model the interactions of Cherenkov photons inside the crystal. Three different thresholds of 3, 6 and 9 p.e. were
considered.} \label{fig:det_eff}
\end{figure}

\begin{figure}[!ht] \centering
\includegraphics[width=0.5\textwidth]{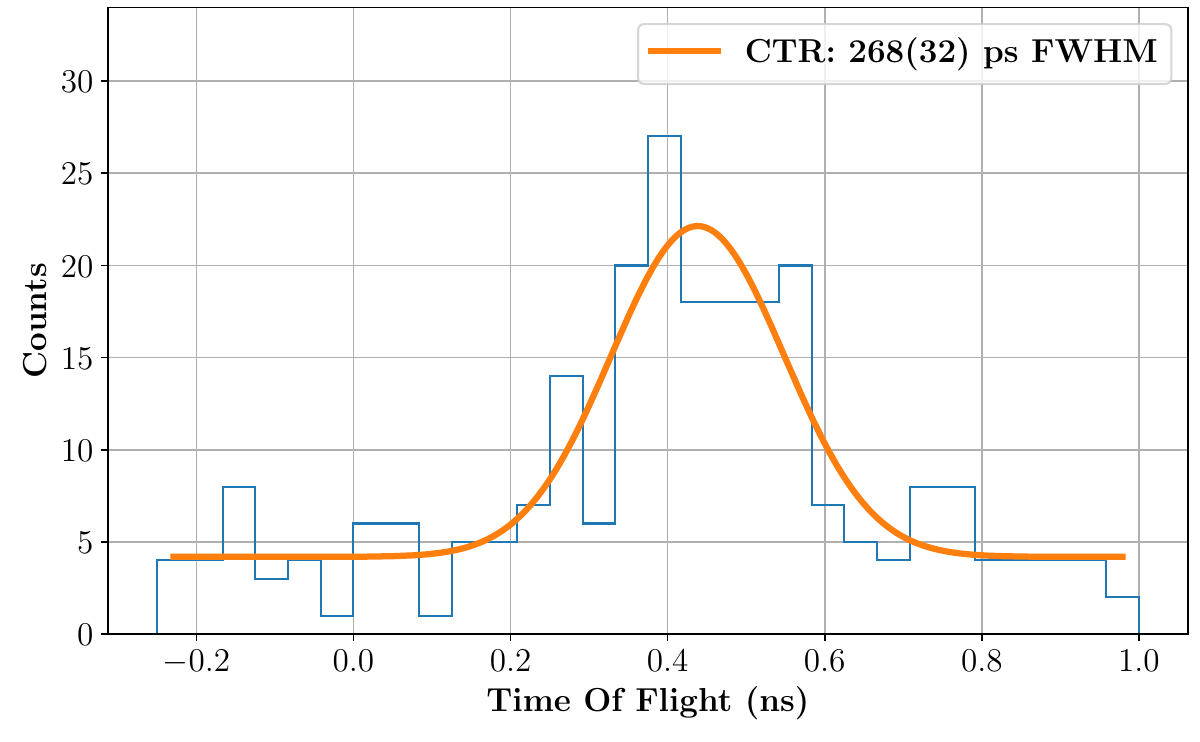} \caption{TOF distribution obtained with the
5 mm thick, 10 cm radius PMMA target irradiated by 63 MeV protons. The PG detector consisted of a
1 cm$^3$ PbF$_2$ crystal coupled to a 3$\times$3 mm$^2$ SiPM and placed at 90$^{\circ}$  
respect to the beam.
Data are fitted with a gaussian distribution convolved with a uniform distribution
of 51 ps width. The resulting FWHM of 268 ps (114 ps rms) corresponds to the gaussian distribution
FWHM and it can be interpreted as the system CTR.} \label{fig:tres}
\end{figure}

\begin{figure}[!ht] 
\centering \includegraphics[width=0.49\textwidth]{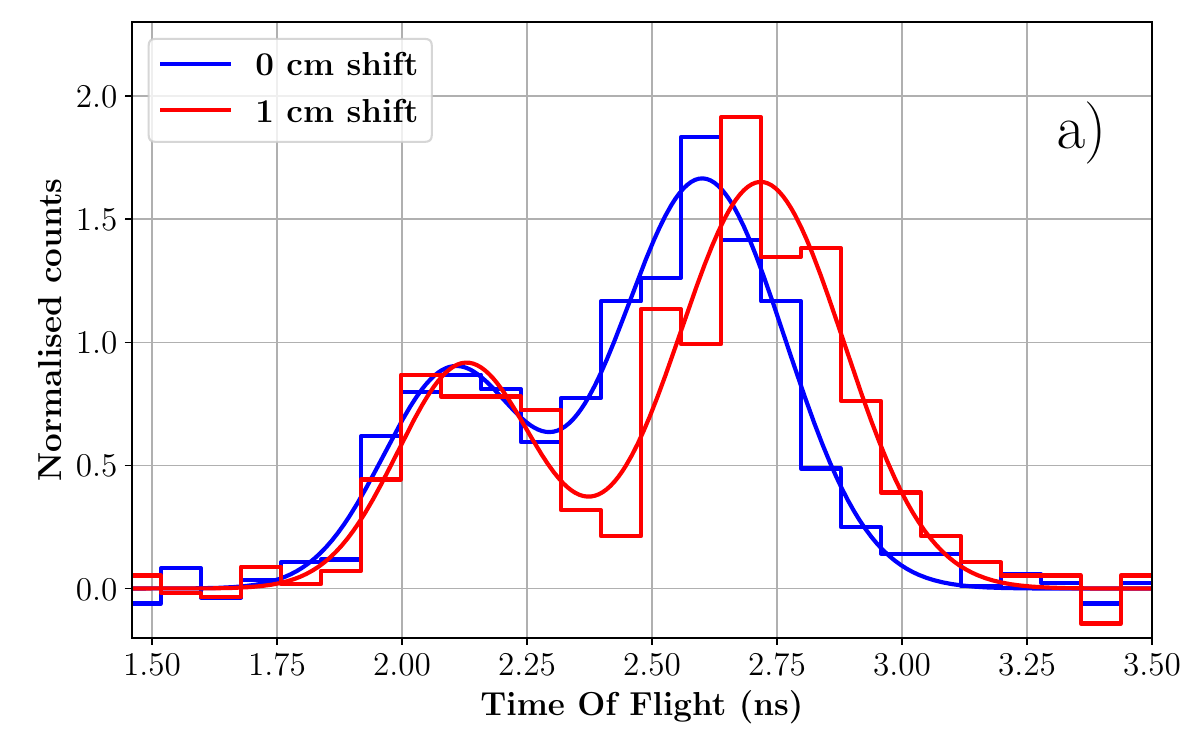} 
\includegraphics[width=0.49\textwidth]{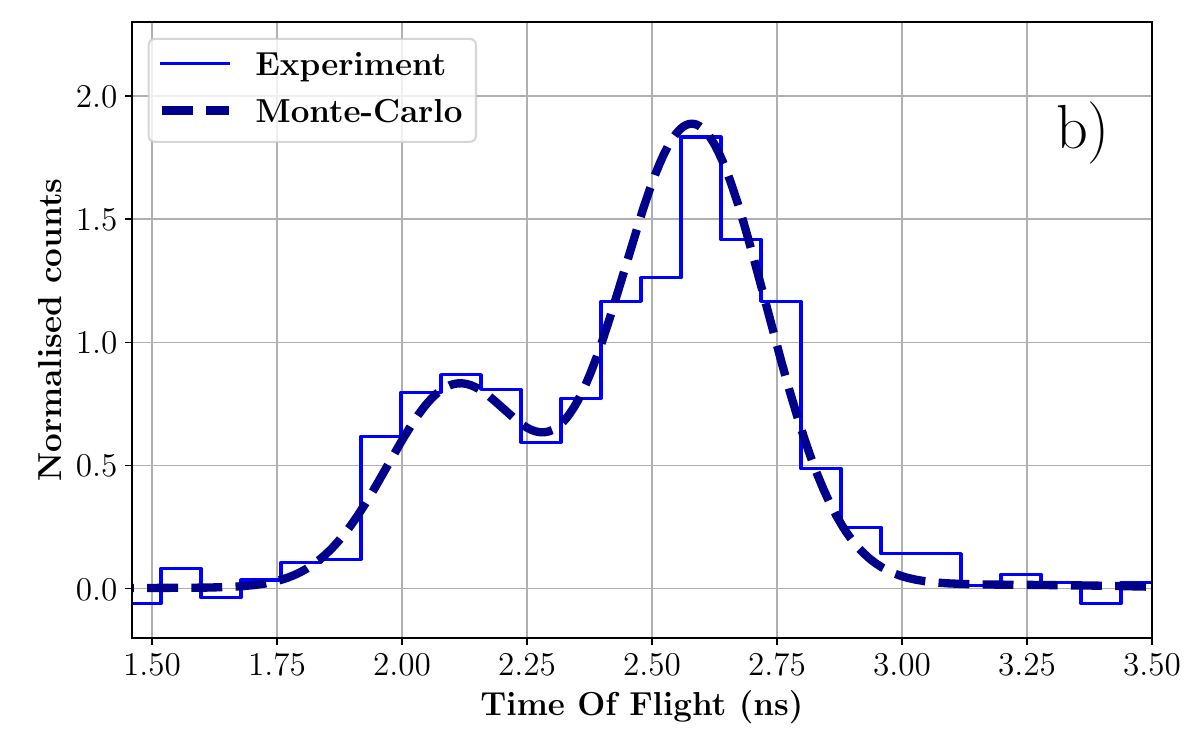}\\
\includegraphics[width=0.49\textwidth]{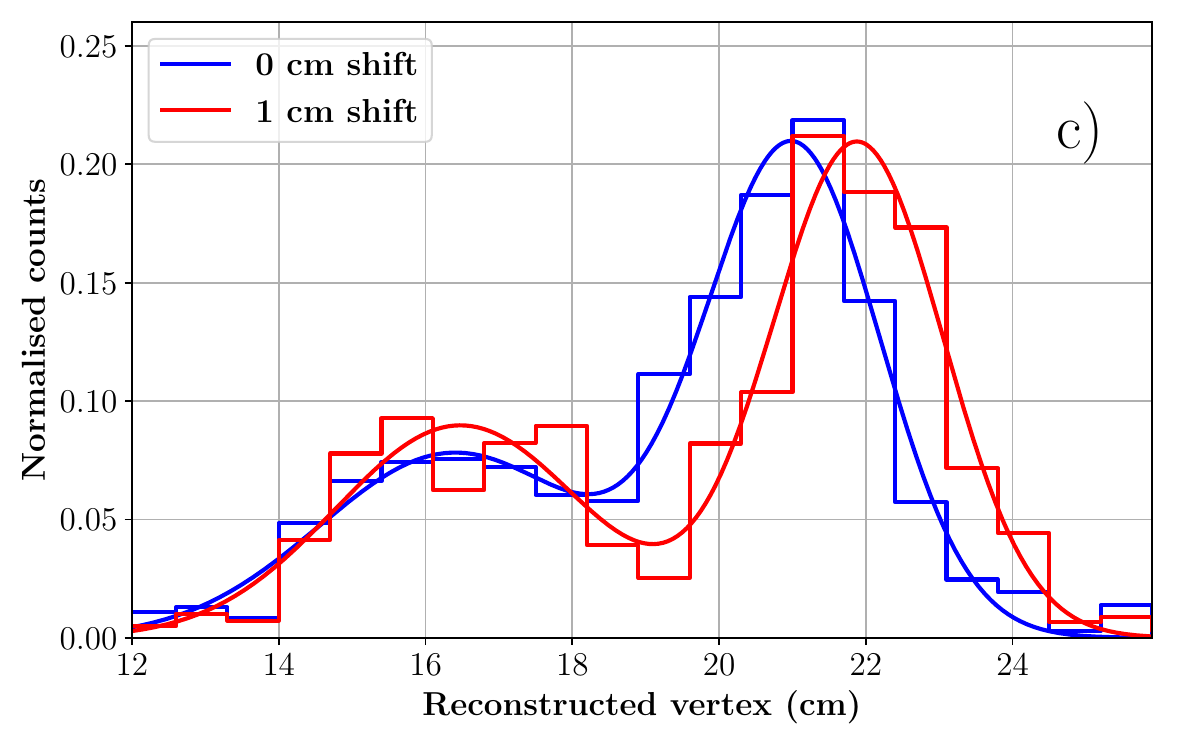} 
\includegraphics[width=0.49\textwidth]{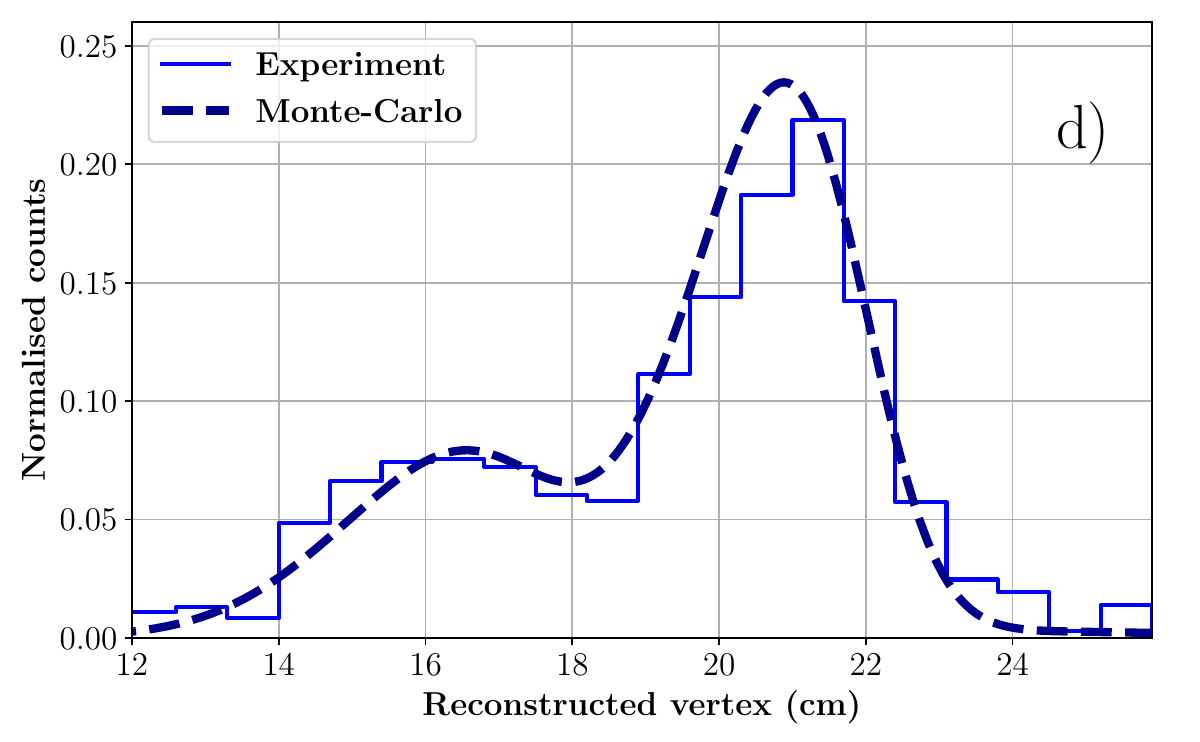} 
\caption{PGT profiles (top) and PGTI
reconstructed profiles (bottom) obtained from the PMMA targets described in figure 
\ref{fig:setup1_63} irradiated with 63 MeV protons.
 In a) the experimental TOF profile obtained for the reference geometry (in blue)
is compared to the one obtained after engendering a proton range shift of 1 cm (in red). The two
histograms are fitted with a double gaussian fuction to improve readability; the fit was not used for
analysis. In b), the simulated
reference profile (dashed line) is compared to the corresponding experimental data. In c) the two
profiles shown in a) are reconstructed with the PGTI algorithm to convert them into the space
domain. In d) the reconstructed simulated (dashed line) and experimental (continuous line) reference
profiles, corresponding to data in figure b) are presented. The experimental profiles are obtained
with a PG module composed of a 1 cm$^3$ PbF$_2$ crystal coupled to a 3$\times$3 mm$^2$ SiPM and placed at 90$^{\circ}$  
respect to the beam. For the 
simulation of the reference profile, we used the Geant4.10.4.p02 toolkit with the QGSP-BIC-EMY
physics list and the UNIFIED model. 
} \label{fig:profiles} 
\end{figure}

\begin{figure}[!ht] \centering 
\includegraphics[width=0.49\textwidth]{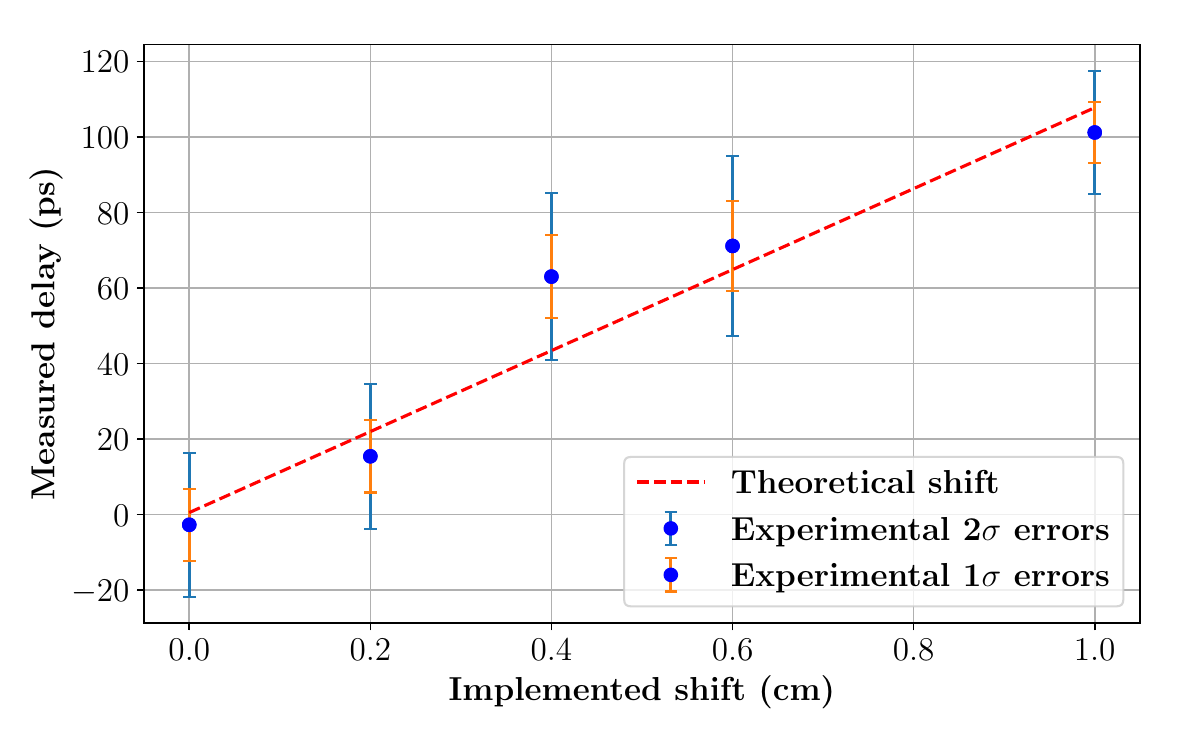}
\includegraphics[width=0.49\textwidth]{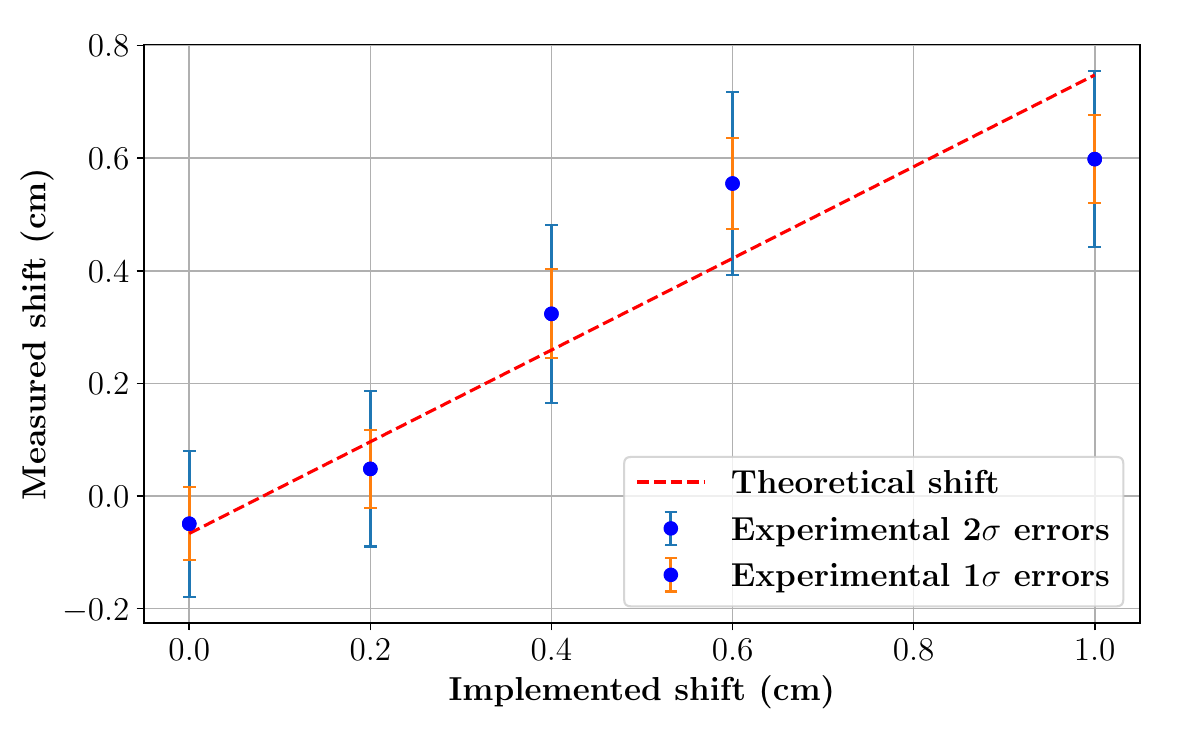} \caption{
Range shift sensitivities obtained with a PMMA target including an air cavity ranging 
from 3 to 4 cm, and irradiated with 63 MeV protons.
On the left, the proton range shift measured with the PGT technique in unit of time is compared to the actual shift implemented in
the phantom. On the right, the PGTI reconstruction allows measuring the proton range shift directly
in the space domain. The dashed red line corresponds to the theoretical correlation between the implemented
shift and the measured parameter as obtained from MC simulation using the Geant4.10.4.p02 toolkit with
 the QGSP-BIC-EMY physics list and the UNIFIED model. Error bars represent the 1$\sigma$ (orange) 
 and 2$\sigma$ (blue) statistical errors obtained by the bootstrap technique (see the Methods section).
 With both techniques, the TIARA detection module
(1 cm$^3$ PbF$_2$ crystal coupled to a 3$\times$3 mm$^2$ SiPM and placed at 90$^{\circ}$  
from the beam direction)  permits to achieve a proton
range shift sensitivity of 4 mm at 2$\sigma$ with a statistics of approximately 600 PGs.}
\label{fig:sensitivity} 
\end{figure}

A detector module based on a 1$\times$1$\times$1 cm$^{3}$ PbF$_2$ Cherenkov radiator coupled to a
3$\times$3 mm$^{2}$  MPPC from Hamamatsu (S13360-3050CS)  read by a commercial preamplifier from
Advansid [https://advansid.com/products/product-detail/asd-ep-eb-pz] was conceived at LPSC
(Laboratoire de Physique Subatomique et Cosmologie). The module was tested at the CAL (Centre
Antoine Lacassagne) MEDICYC facility \cite{hofverberg_60_2022}  with protons of 63 MeV delivered 
in bunches of 4 ns duration every 40 ns period. \\ A first cylindrical PMMA target (density = 1.19 g/cm$^3$)
of 0.5 cm (or 1 cm) thickness and 10 cm radius (cf. figure \ref{fig:setup1_63}) was employed to
characterise the module energy and time response, followed by a cylindrical PMMA block of 23 cm thickness 
(enough to stop the beam) and 10 cm radius, placed at 3 cm distance from the thinner PMMA target.
 A single crystal (sc) diamond detector of
$4.5\times4.5\times0.55$ mm$^3$ volume from Element6 [https://e6cvd.com/] was used as a beam monitor
to tag in time the incident protons.  The diamond was read-out on both faces using a commercial C2
preamplifier from Cividec [https://cividec.at/electronics-C2.html]; the two signals were summed up
for analysis in order to increase the SNR and therefore improve the time resolution as proposed by
Curtoni et al.\cite{curtoni_performance_2021}. A 8 mm thick, 2 mm diameter brass collimator was used
to match the beam size to the limited diamond detection surface. 
The effective (after collimation) beam intensity reaching the beam monitor was estimated \textit{a posteriori} 
to amout $\sim$0.025 p/bunch from Poisson statistic considerations (see the Methods section).
A time resolution of 156 ps (FWHM) was previously measured for sc-diamonds in the same energy range
 \cite{marcatili_ultra-fast_2020} and similar experimental conditions. 
The gamma detector module was positioned on the side of the
large PMMA target at 14 cm  from the beam axis and facing the Bragg-peak region in the thick
target (orthogonal to the beam). All signals were acquired in time coincidence with the beam monitor  and then digitally
sampled using a HDO6104A-MS LeCroy oscilloscope  with 1GHz bandwidth, 10Gs/s and a 12 bit ADC. The
analysis was performed offline. In order to obtain a perfect SPR, the small fraction of 2-protons events
acquired was cut-off during the analysis.\\ \\
%
\noindent \textbf{Energy response.}~Signals from  the TIARA module were integrated to record the
detector energy response as shown in figure \ref{fig:energy}, left. On the right, the simulated
energy distribution of PGs from 63 MeV protons impinging on a PMMA target is shown for comparison.
The two spectra are clearly unrelated as the TIARA module does not allow to establish the PG
incident energy. As a result, none of the intense, characteristic PG emission lines visible in the
right plot can be distinguished in the left plot. Actually, because of the limited detection
volume,  the gamma ray energy is not fully deposited in the detector. As a consequence, multiple
interactions of mono-energetic gamma rays  may result in a  wide range of deposited energies. This
phenomenon, combined to the relatively low light yield of Cherenkov radiators in the 2$-$10 MeV
range, results in the typical SiPM single photo-electron (p.e.) spectrum shown in figure
\ref{fig:energy} (left), in which single Cherenkov photons are literally counted by the device. \\
An acquisition threshold of 6
p.e. was applied to the gamma detector signal in order to reduce the probability of triggering on 
SiPM dark counts. For the same purpose, the coincidence with the beam monitor was also imposed for data acquisition. 
A median number ($N$) of 7 p.e. was detected, corresponding to the amount of Cherenkov photons
detected by the SiPM. The value of $N$ indirectly affects the module time resolution as the SiPM
contribution  roughly goes as $SPTR/\sqrt{N}$ \cite{gundacker_high-frequency_2019}, where $SPTR$ is
the intrinsic Single Photon Time Resolution (SPTR) of the SiPM. \\ 
The poor proportionality between incident and deposited PG energy mainly affects  
the determination of the acquisition threshold as, at low energies, the module detection efficiency dramatically  depends on the PG incident energy. This can be estimated through 
 MC simulation (see the Methods section). Neglecting the geometrical contribution, the intrinsic
 detection efficiency of a PG module
 as the one described in figure \ref{fig:setup1_63} depends on two contributions: 
 the probability for the PG to interact
in the crystal and the probability that this interaction produces more than $N_{th}$ photoelectrons,
with $N_{th}$ being the threshold in p.e. applied for data acquisition. Figure \ref{fig:det_eff} shows the detection efficiencies as a function of PG energy obtained by MC simulation for three different thresholds of 3, 6 and 9 p.e. It can be observed that the threshold  does not provide a sharp energy cut-off. For example, using a threshold of 6 p.e. gives a 5\% probability of detecting PGs with an energy of 2 MeV. \\ 
We will show in the next paragraphs that the lack of energy resolution does not
compromise the proton range measurement sensitivity when a very high time resolution is achieved
as long as the detection efficiency is properly taken into account. \\
\\
\noindent \textbf{Coincidence Time Resolution.}~The time stamps of both the diamond detector and the
TIARA detection module have been recorded to obtain the proton plus gamma $TOF$ variable in equation
\ref{eq:pgti}. The two diamond signals (one from each face) are first summed-up to increase their
SNR, then a digital Constant Fraction Discrimination (CFD) with a  50\% CF value is applied to both
the PG and the diamond detectors. The resulting PG TOF distribution is presented in figure
\ref{fig:tres} for events corresponding to single protons in the beam monitor.
This distribution is a measure of the system Coincidence Time Resolution (CTR): more precisely, its
width results from the convolution of the system intrinsic time resolution and the proton transit
time in the 5 mm PMMA target. The latter can be approximated by a uniform distribution of $\sim$51
ps width (from Geant4 simulations). After deconvolution of the proton transit time, and the
suppression of the flat background, the best system time resolution is estimated to be 268 $\pm$ 32
ps (FWHM) for data in figure \ref{fig:tres}. The same analysis was carried out independently  on
three  TOF distributions acquired with either the 5 mm or the 1 cm thick targets; the averaged time
resolution obtained from these repeated experiences is of 315 $\pm$ 40 ps (FWHM). The CTR can also
be interpreted as the quadratic sum of the beam monitor's and the PG detector's time resolutions
under the hypothesis that the two contributions are independent and gaussians. Since the diamond
time resolution was measured to be 156 ps (FWHM) in similar
conditions\cite{marcatili_ultra-fast_2020}, the TIARA detection module time resolution can be
estimated to approximately amount 276 ps (FWHM). Therefore, for this detection module design and for
63 MeV protons, the system CTR is dominated by the PG time measurement. \\ 
\\ 
\textbf{Study of PGT sensitivity.}~The 1 cm thick PMMA target was placed downstream the 
diamond detector (cf. figure
\ref{fig:setup1_63}) with the thick target positioned at 3 cm distance. This configuration
(considered as the reference geometry)  was employed to simulate an air cavity heterogeneity in a
uniform anatomy. The distance between the two targets was then progressively increased by 2, 4, 6
and 10 mm to reproduce an unpredicted variation in the anatomy and consequently a shift in the
proton range. For each shift (0, 2, 4, 6 and 10 mm), the TOF distribution between the proton beam
monitor and the PG detector was recorded. Figure \ref{fig:profiles}a shows, as an example, the TOF
histograms measured at 3 cm  (0 mm shift) and 4 cm (10 mm shift) target-to-target distance. 
The two distributions are clearly separated: the distal fall-off of the 4 cm shift distribution is displaced
towards higher TOFs according to the air cavity thickness introduced between the targets. The flat
background contribution, due to the random coincidences between  SiPM dark counts and protons
interacting in the diamond, is subtracted before  analysis, and only 1-proton signals in the beam 
monitor are considered. After  background rejection, each of the
5 TOF distributions includes around 600 PG events. For this experiment,
 it is not possible to  establish the number of protons delivered as the scope dead
 time when writing the waveforms' values on disk considerably affects the total acquisition time. 
 Nevertheless, previous MC 
simulations\cite{jacquet_time--flight-based_2021} allow to estimate that 600 PG events 
would correspond approximately to 2~$\times$10$^{6}$ incident protons for the full 30-channels TIARA prototype. 
\\ 
The  dashed curve in figure
\ref{fig:profiles}b represents the simulated PG profile in the reference geometry (0 mm shift), that was
obtained taking into account the energy dependency of the  PG module efficiency presented in figure
\ref{fig:det_eff} (see the Methods section). This
curve is used as a term of comparison to perform relative measurements of the proton range shift 
according to the approach previously described\cite{marcatili_ultra-fast_2020,
jacquet_time--flight-based_2021} and summarised in the methods section. This methodology measures the 
proton range shift only exploiting the distal region of the PG profiles without making any assumption on the profiles' shape. The double gaussian fit
(continuous lines in figure \ref{fig:profiles}, left) are only presented as  eye-guide but  
were not used for the analysis.
In the right plot, an excellent
agreement between the experimental and the simulated reference profiles can be observed, serving as
a validation of the detector model developed in Geant4.\\ 
The measured shifts are reported in figure \ref{fig:sensitivity}, left, as blue experimental data
points, as a function of the implemented target shift. The dashed red line represents the
correlation between the measured time delay and the shift implemented in the geometry as obtained by
MC simulation. Its slope amounts to 107 ps/cm, roughly corresponding to the proton speed at the exit
of the thin target (the simulated value amounts to 109 ps/cm). The orange and blue error bars
respectively summarise the 1$\sigma$  and 2$\sigma$ experimental errors obtained by bootstrapping
methods (see methods section). The experimental errors on these data indicate that the TIARA
detection module is able to measure a 4 mm proton range shift at 2$\sigma$ confidence level, from
the exclusive measurement of TOF  with a very low statistics of acquired PGs. The same data also
show that a proton range sensitivity of 2 mm would be achievable at 1$\sigma$ confidence level.
\\ \\
\textbf{Study of PGTI sensitivity.}~The five TOF datasets acquired for the different target shifts
are reconstructed  on an event-by-event basis according to equation \ref{eq:pgti} and following the
methodology described in \cite{jacquet_time--flight-based_2021}. With this procedure, the TOF
distributions in figure \ref{fig:profiles}a are converted into PG profiles in the space domain
providing straightforward means to directly measure the range shift in mm instead of ps (see figure
\ref{fig:profiles}c). In analogy to figure \ref{fig:profiles}a, data are shown for the 3 and 4 cm
air cavities. The time-to-space conversion is necessary to combine the response of multiple modules
placed at different angular coordinates. Actually, the PG TOF depends on the relative position
between the PG vertex and the PG detector and it must be deconvolved before summing-up TOF
distributions obtained with different modules. In this work, only one TIARA detection module was
available: our goal  therefore was to demonstrate that PGTI can provide the same sensitivity than
PGT. \\ The proton range shift was evaluated with respect to the simulated PGTI profile in reference
conditions (dashed line in figure \ref{fig:profiles}d) applying the same approach used for PGT
distributions after background subtraction (see methods section). In analogy with PGT analysis, the
measured proton range shift is reported in figure \ref{fig:sensitivity}, right, against the applied 
cavity shift. The 1$\sigma$ and 2$\sigma$ statistical errors are also shown as orange and blue error
bars, respectively. The same proton range sensitivity is observed for PGTI and PGT: a 4 mm range
shift could be distinguished at 2$\sigma$ confidence level. Nevertheless, an offset of 0.66~mm is
present on the PGTI dataset, as highlighted by the dashed red line. This curve represents the
expected correlation between the measured and the implemented range shift as obtained by MC
simulations; here its intercept has been adjusted to the data points to properly estimate the range
offset. This offset results from the propagation, during reconstruction, of the systematic errors
made when measuring the coordinates of the TIARA detection module and those of the beam monitor. In
the future, this error can be easily minimised by imaging the target (or the patient) and the
experimental set-up with a CT scanner in order to establish the detectors' positions  with
sub-millimetric precision during the treatment planning phase. Despite this offset, the behaviour of
the experimental points in figure \ref{fig:sensitivity}, right,  is still linear with a slope of
0.81, a value inferior to one as the current reconstruction is biased by the $T_{p}$ term (cf.
equation \ref{eq:pgti}) determined in reference conditions. This effect could be avoided with an
iterative reconstruction approach, as the one proposed by Pennazio et al.\cite{pennazio_2022},  that could 
establish offline the actual range shift. Nevertheless,
in this work we focussed on an event-by-event reconstruction that can be implemented online during
data acquisition, hence allowing to promptly stop the treatment  if a significant shift is detected.
To achieve this goal, we do not need the actual value of the proton range shift, but only to
establish whether a statistically significant discrepancy with respect to the treatment plan exists
or not.\\
\subsection*{Detector characterisation with 148 MeV protons } 
\begin{figure}[!ht] \centering \includegraphics[width=0.49\textwidth]{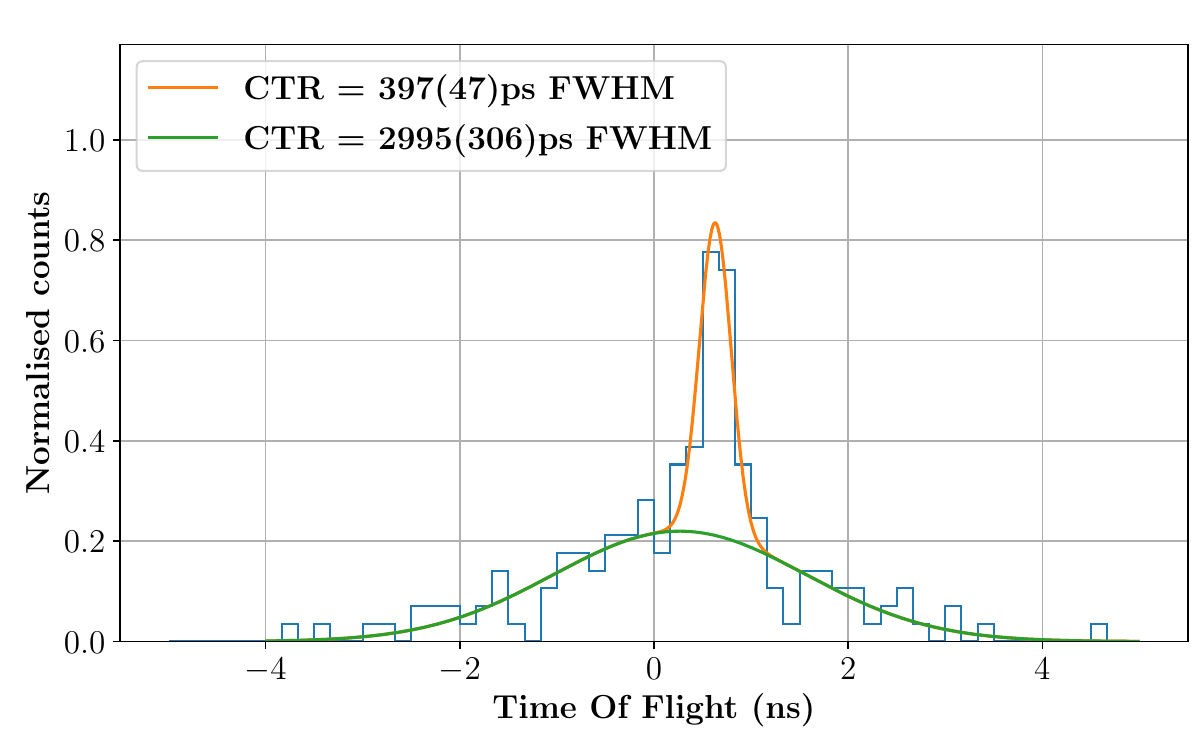}
\includegraphics[width=0.49\textwidth]{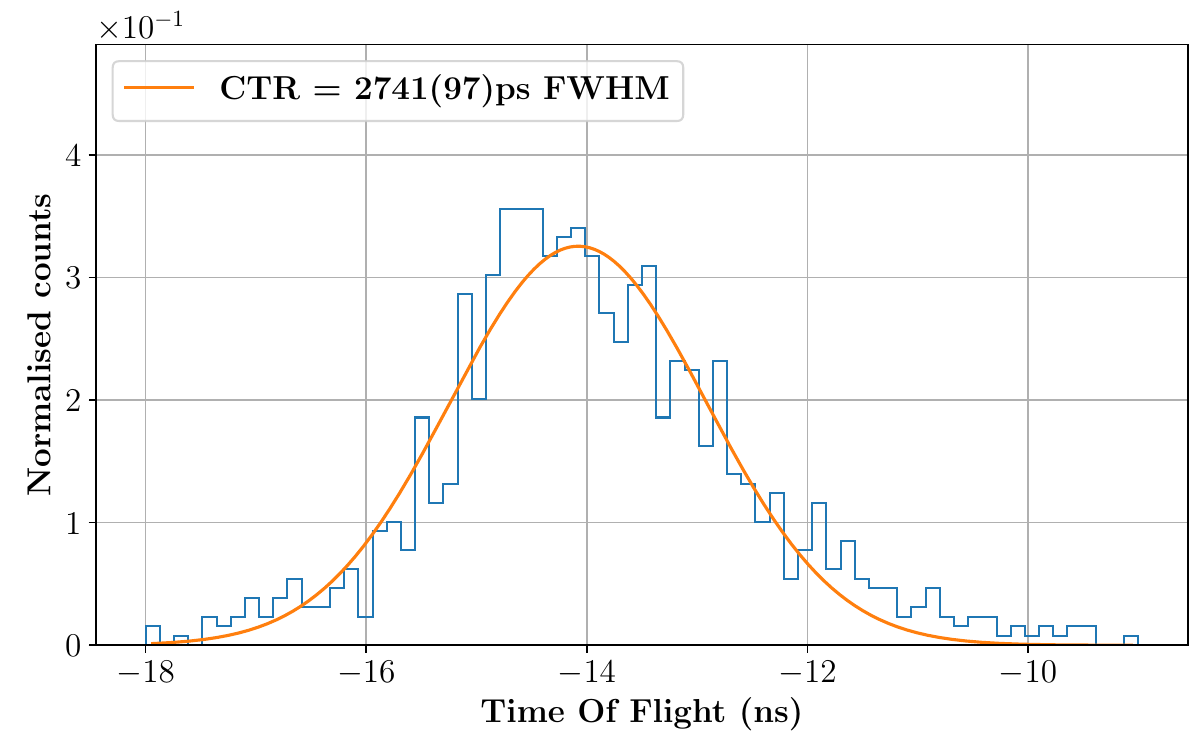} \caption{On the left, TOF distribution
obtained from the irradiation of the 5 mm target. The gaussian peak with 397 ps (FWHM) corresponds
to PGs generated by protons traversing the beam monitor; the large background signal corresponds to
PGs generated  by protons passing by the diamond detector. On the right, the TOF distribution
obtained from random coincidences between the diamond and the PG detectors  confirms the origin of
the background on the left plot:  it is a measure of the bunch-induced time resolution.  }
\label{fig:p1_thintarget} \end{figure}

\begin{figure}[!ht] \centering \includegraphics[width=0.5\textwidth]{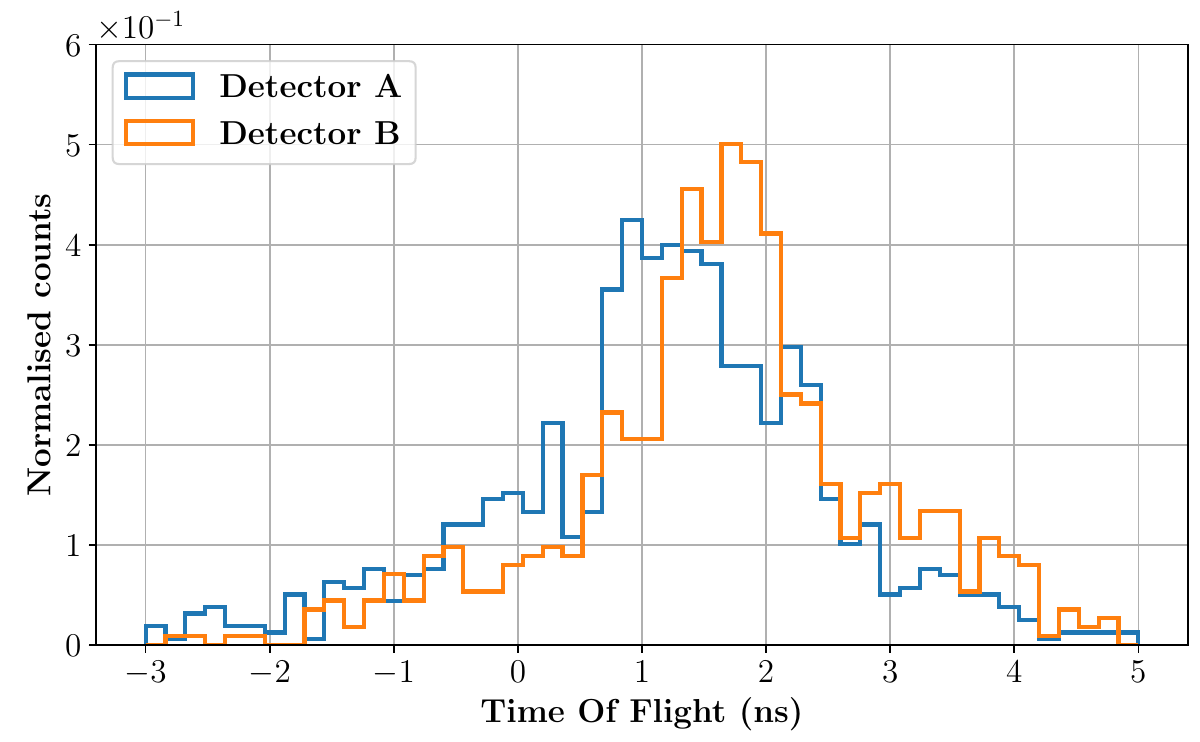} \caption{TOF
distributions obtained with the thick PMMA target and with detector A (in blue) placed at 157$^{\circ}$ and 
detector  B (in red) placed at 90$^{\circ}$ from the beam direction. The relevant
signal is in the region  between 0.5 and 3 ns. Outside this region, the background is mainly due to
random coincidences caused by the limited size of the beam monitor. The two bumps located at
($-$1.0~$\div$~0.5) ns and (2.9~$\div$~4.1) ns for detectors A and B distributions, respectively, 
are associated to  protons
scattered in the beam detector.} \label{fig:2spectra}
\end{figure}

A second version of the detector
module was realised to improve the  detection efficiency without compromising the time resolution. 
It is composed of a 1$\times$1$\times$2 cm$^3$ PbF$_2$ crystal coupled to a 6$\times$6  mm$^2$ MPPC 
from Hamamatsu (S13360-6075CS) and read by a custom preamplifier based on the design of Cates et al.
\cite{cates_improved_2018}. The module was tested with 148 MeV protons at the ProteusOne S2C2 synchrocyclotron 
at CAL. S2C2 delivers protons in micro-bunches of 16~ns period with a 50\% duty cycle
\cite{de_walle_diamond_2016}: this micro-structure is embedded in a macrostructure characterised 
by 8 $\mu$s pulses every ms. In this paper, the 8 $\mu$s beam structure will be referred to as proton pulse, whereas the 16~ns micro-structure will be mentioned as proton bunch. The beam profile is Gaussian with a measured standard deviation of 4.3
mm at 148 MeV\cite{pidikiti_commissioning_nodate}. 
The same sc-diamond detector used in the 63 MeV experiment, read on both sides using Cividec C2
preamplifiers, provided the time stamps for the incident protons, 
The effective intensity at the beam monitor level was arbitrarily set to a low value that was estimated
\textit{a posteriori} to amount $\sim$0.78 p/bunch on average. \\
\begin{figure}[!ht] \centering \includegraphics[width=0.49\textwidth]{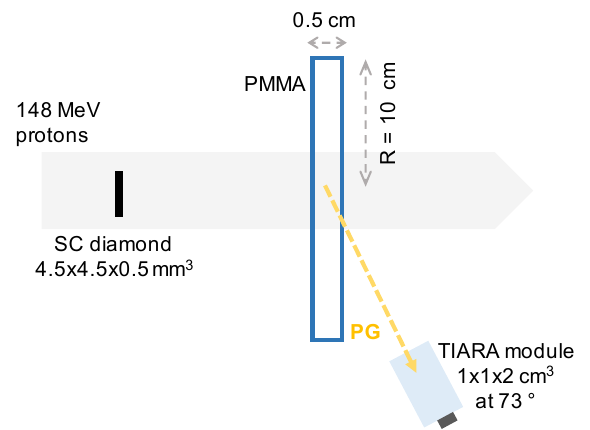}
\includegraphics[width=0.49\textwidth]{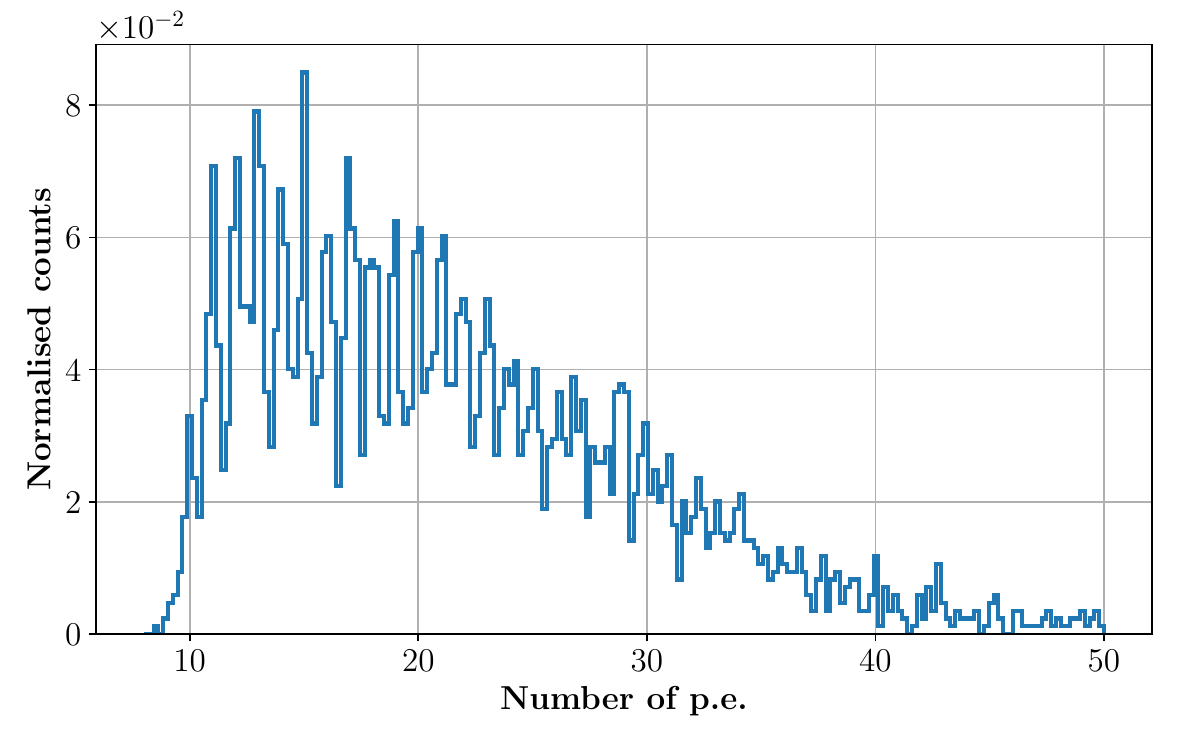} \caption{Left: experimental set-up for
the CTR measurement carried out at the S2C2. A 5 mm thick, 10 cm radius PMMA target was used as
a source of PGs to measure the CTR. Given the limited transit time of protons within the target, the
PG source can be considered point-like. The beam size (4.3 mm $\sigma$) was larger  than the diamond
surface (4.5$\times$4.5~mm$^2$) as no collimator was employed. The TIARA detection module was
composed of a 2 cm$^3$ PbF$_2$ crystal coupled to a 6$\times$6 mm$^2$ SiPM and placed at approximately
73$^{\circ}$  from the beam direction. The effective beam intensity at the beam monitor
level was 0.78 p/bunch. Right: energy deposited
in the TIARA detection module expressed as the integral of the SiPM signal. A threshold of 10 p.e. 
was applied for data acquisition, resulting in a median number of 21 p.e. detected.} \label{fig:setupp1_1}
\end{figure}
In a first experiment (cf. figure \ref{fig:setupp1_1}, left), a 5~mm thick, 10 cm radius cylindrical PMMA
target (density~=~1.19~g/cm$^3$) was employed
in a configuration similar to the one used at the MEDICYC accelerator in order to measure the system
CTR. 
The TIARA module was placed close to the target axis in order to increase the geometrical efficiency,
 at approximately 73$^{\circ}$. Signals from
all detectors were digitally sampled using a Wavecatcher module \cite{breton_wavecatcher_2014} with
500 MHz bandwidth and a sampling rate  of 3.2~Gs/s. 
 The acquisition was triggered by the coincidence of the two detectors within a 15 ns time window. 
 The analysis was performed offline. The SPR was 
ensured by selecting only 1-proton events in the diamond detector. Scattered protons directly
detected by the SiPM in the PG module were rejected by pulse-shape analysis as they produce longer
signals than those associated to Cherenkov events in the crystal.\\
The energy response of the module is shown in figure \ref{fig:setupp1_1}, right: a median
number of 21 p.e. is detected per each PG event when applying an acquisition threshold of 100 mV
($\sim$10 p.e.). This improvement, with respect to the median value of 7 p.e. obtained in the
previous experiment (figure \ref{fig:energy}, left) is mainly due to the increased size of the SiPM
that allows collecting a  larger number of Cherenkov photons per PG event. 
The difference between the PG detector time stamp (obtained with a 5\% CF threshold) and the diamond
time stamp (obtained with a 50\% CF threshold) was calculated to build the TOF distribution for the
5~mm target irradiation shown in figure \ref{fig:p1_thintarget}. This distribution presents two
clear components: i) a well resolved gaussian peak (highlighted by the orange fit) that corresponds
to PG events generated by protons crossing the diamond detector, and ii) a broad background
(highlighted by the green fit) associated to PG events from protons passing by the beam monitor. \\
The first peak directly provides a measurement of the system CTR with a value of 397 ps (FWHM). This
CTR can also be interpreted as the quadratic sum of the diamond  and the PG detector time
resolutions. In a previous experiment, the intrinsic time resolution of the diamond was measured to
be 360 ps (FWHM) when reading-out a single face of the detector. Under these conditions, the diamond
time resolution is expected to be overestimated by 50$-$70 ps (FWHM) with respect to experiments in
which signals from both sides are summed-up for analysis \cite{curtoni_performance_2021}. Despite
the measured time resolution can  not be directly compared to the CTR measured in this work,  it
still appears obvious that, for 148 MeV, the system CTR is now dominated by the beam monitor, 
with the TIARA detection module contributing for less than 167 ps (FWHM). This effect
is expected, as an increase in the proton energy corresponds to a decrease in the energy deposited
in the diamond: $\sim$0.9 MeV is deposited in a 0.55 mm thick diamond by 148 MeV protons instead of
$\sim$1.7 MeV for 63 MeV protons. \\ 
The background component in the TOF distribution of figure
\ref{fig:p1_thintarget} is a direct effect of the limited size of the beam monitor. Since no
collimator was used in this experiment, only a fraction of the beam (approximately 20\%) traverses
the diamond detector. This means that, while the effective beam intensity was estimated to 0.78 p/bunch
at the beam monitor level, the actual beam intensity is of the order of 4.7 p/bunch. 
Therefore the acquisition of part of the events may be triggered by the
coincidence of a PG (or another secondary particle) with a random proton  that has not produced 
the gamma ray (or the secondary particle). In other words, the 
background distribution represents the random coincidences between the PG detector and the beam
monitor: its shape is not flat, because the beam time-structure is periodic with a nominal bunch
width of 8 ns and a nominal standard deviation of $8/\sqrt{12}$ ns. The background could be fitted
with a Gaussian distribution of 1.27 $\pm$ 0.13 ns sigma, a value that is consistent with the S2C2 micro-bunch
standard deviation. This hypothesis is confirmed by a separate, more direct measurement of the bunch
width. A random coincidences TOF distribution was built between events from the PG module and the
beam monitor triggering on the bunch arriving immediately before the 15 ns coincidence window. The
distribution obtained, shown in figure \ref{fig:p1_thintarget}, right, is compatible with  the
background  in figure \ref{fig:p1_thintarget}, left, with its mean value shifted by the  micro-bunch
period (16~ns). The standard deviation of this distribution is 1.17 $\pm$ 0.04 ns, in agreement with
the measurement in figure \ref{fig:p1_thintarget} left.\\
\begin{figure}[!ht] \centering \includegraphics[width=0.45\textwidth]{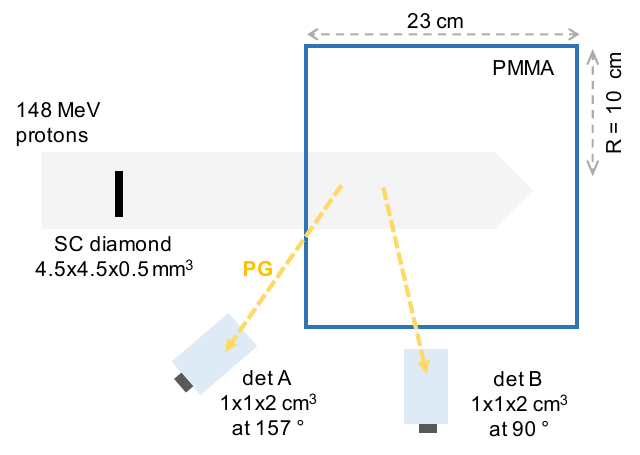}
\hspace{0.3cm} \includegraphics[width=0.4\textwidth]{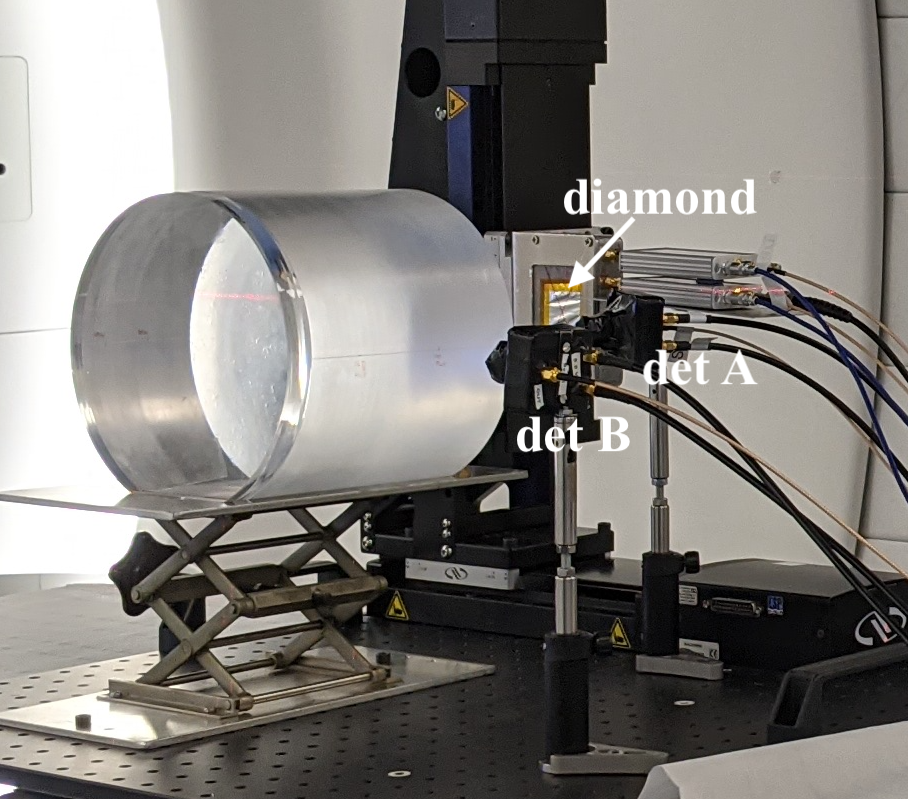} \caption{Experimental
set-up for the PGT profile measurement at the S2C2 facility. The 23 cm thick, 10 cm radius PMMA target
fully stopped the 148 MeV protons after a range of 13.4 cm. Two gamma detector modules were placed
upstream the target (det A, at 157$^{\circ}$) and at the Bragg peak (det B, at 90$^{\circ}$). 
Each module was composed of a 2 cm$^3$
PbF$_2$ crystal coupled to a 6$\times$6 mm$^2$ SiPM. The effective beam intensity at the beam monitor
level was 0.78 p/bunch.} \label{fig:setup2} \end{figure}
In a second experiment, the thin target was replaced by a 23 cm thick, 10 cm radius PMMA target to stop 148 MeV
protons after a 13.4~cm range. Two identical gamma detectors were placed close to the beam entrance
(position A, at 157$^{\circ}$), and close to the Bragg peak (position B, at 90$^{\circ}$) 
both aiming at the Bragg peak region as shown in figure \ref{fig:setup2}.
During analysis, only 1-proton events were considered and scattered protons directly
detected by the SiPM in the PG module were rejected by pulse-shape analysis.
The TOF distributions measured for detectors A and B are presented in figure \ref{fig:2spectra}. The
same effect described in figure \ref{fig:p1_thintarget} is visible here: because of the beam monitor
limited size, a background associated to random coincidences is recognisable in both distributions.
A second background component is present in the TOF ranges ($-$1.0~$\div$~0.5) ns and
(2.9~$\div$~4.1) ns for detectors A and B distribution, respectively. These events are associated to
scattered protons generating PGs in the gamma detector (most probably in the packaging) 
and will be discussed in detail in the next section. Nonetheless, the PG signal is still clearly 
detected between 0.5 and 3 ns for both
detectors. Detector A is directed at the beam entrance while detector B is closer to the Bragg peak
region. Thus, detector A has a larger solid angle for the measurement of PGs at the target entrance
with a reduced efficiency for the Bragg peak region, whereas the opposite is true for detector B,
explaining the different positions of the PG profiles maxima and their shapes. Traditionally, PGT
detectors are placed at beam entrance to take advantage of the overall longer TOF, thus
increasing the TOF measurement sensitivity \cite{golnik_range_2014, werner_processing_2019}, and to avoid
scattered protons that are mostly forward-directed.
However, this approach limits the PG statistics acquired in the fall-off region where the proton
range measurement is performed \cite{hueso-gonzalez_first_2015} and it ultimately reduces the
technique sensitivity; while the profile fall-off is very sharp for detector B, the one obtained
with detector A is not well defined. Combining the readings of multiple detectors placed at
different angular positions around the target is therefore the only way to achieve a uniform
statistical efficiency and a uniform sensitivity throughout the whole proton range. This possibility
could lead to further exploit the measured PG profile in order to assess anatomy variations in
patients \cite{polf_prompt_2009}, but it requires a dedicated reconstruction as the one proposed by
PGTI in order to appropriately sum-up the response of multiple detectors. As a tangible example, TOF
distributions acquired with detectors in position A and B are non-uniformly shifted and stretched in
the time domain, according to the variation of the  PG TOF all over the proton range. While the
proton transit time in the target is, by definition, the same for the two detectors, the PG TOF
varies from 165 to 507 ps for detector A (342 ps of relative PG delay), and from 329  to 407 ps for
detector B (78 ps of relative PG delay) for PGs generated at the target entrance and for those
generated at the Bragg peak respectively: the two  TOF distributions cannot be summed-up in the time
domain without causing blurring and a loss of resolution/sensitivity. Alas, in this experiment, the
presence of the broad background from random coincidences in the beam monitor prevented PGTI data
reconstruction. In fact, the current PGTI algorithm needs the system time resolution as an input
parameter for the reconstruction \cite{jacquet_time--flight-based_2021}, and in presence of a double
component it is impossible, for a single event, to establish the associated time resolution. Our
current effort is in the development of a larger area beam monitor in order to cover the whole beam surface
\cite{gallin-martel_large_2018}. 
%
\subsection*{Background events in Cherenkov detectors} 
\begin{figure}[!ht] \centering \includegraphics[width=0.5\textwidth]{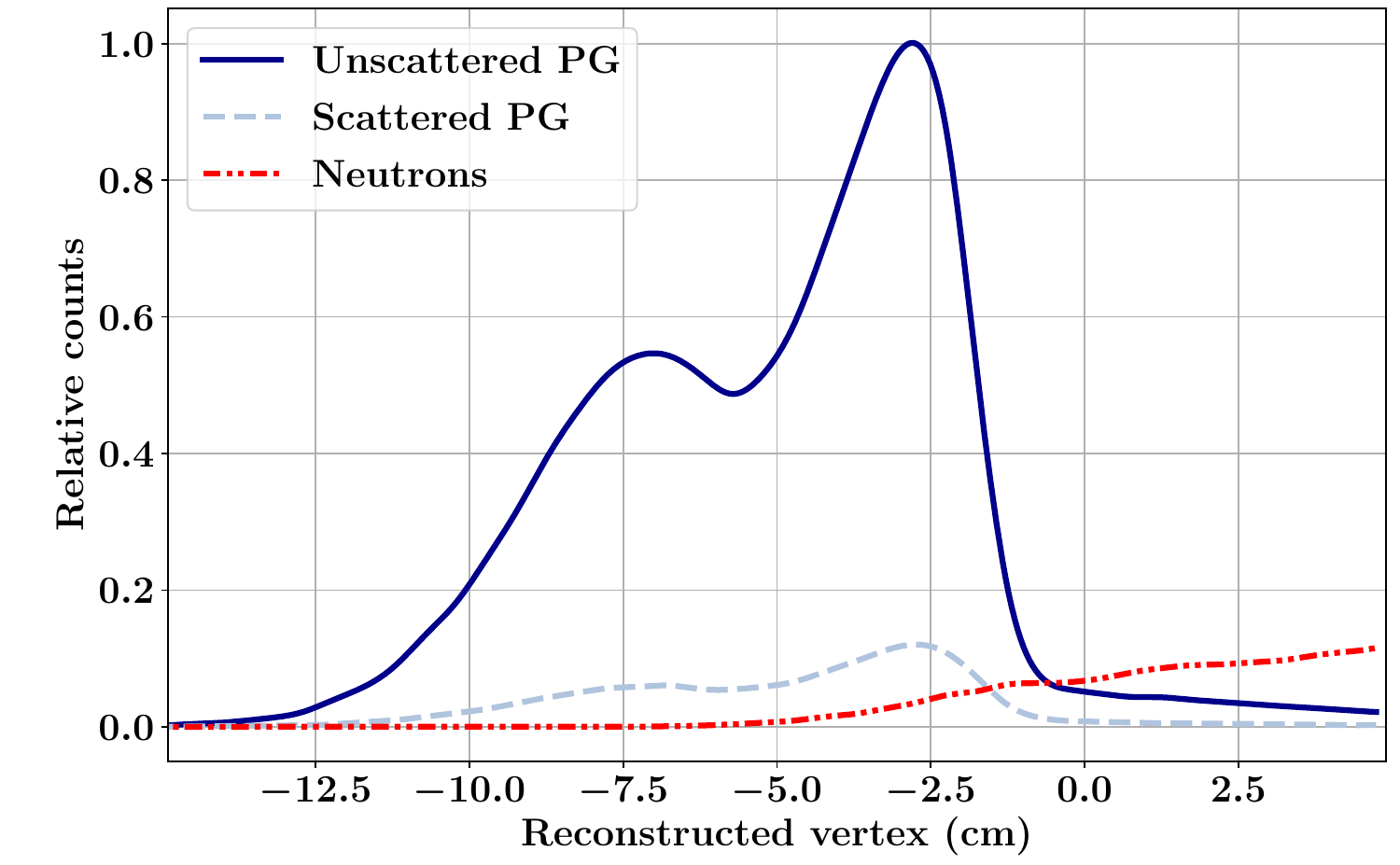}
\caption{Vertex distributions of different secondary particles generated by a 100 MeV proton beam
impinging on the spherical phantom head described in Jacquet et 
al.\cite{jacquet_time--flight-based_2021}. 
 Data are obtained by MC simulation (Geant4.10.4.p02 toolkit with 
the QGSP-BIC-EMY physics list); the detector is not simulated. The contribution of PG scattered in the phantom is reported
separately. It can be observed that their profile has the same shape as the one from unscattered PGs and
therefore they constitute a valuable signal.
The contribution of  secondary neutrons (in red) is not constant and
cannot be easily rejected by TOF without compromising the measurement of the PG profile fall-off.
Data are taken from Jacquet et al.\cite{jacquet_time--flight-based_2021}.} \label{fig:neutrons}
\end{figure}

One of the hypotheses that has motivated the
use of Cherenkov radiators for PG imaging is their insensitivity to neutrons. 
A PG detector is
subjected to three main sources of background noise: neutron and neutron-induced gamma-rays that are
time-correlated to the proton beam (i.e. originating in the nozzle or in the patient); neutron
and neutron-induced gamma-rays from the environment; and protons scattered in the 
patient/target and, in our case, the beam monitor.\\
From MC simulation
\cite{jacquet_time--flight-based_2021}, it has  already been demonstrated that, for most of the
particles falling in the first category, the contribution is not flat, but it increases at the
fall-off of the PG profile (see neutron contribution in figure \ref{fig:neutrons}, data from
\cite{jacquet_time--flight-based_2021}). Thus, even if this component may be rejected by TOF
selection \cite{marcatili_ultra-fast_2020}, still, the fall-off of the PG profile will be biased,
with a direct consequence on the accuracy of the proton range measurement. \\
Conversely, non
time-correlated background particles result by definition in a constant baseline in the TOF
distribution. The absence of a collimator is pivotal in keeping the level of this environment noise
negligible. The exploitation of a threshold process such as Cherenkov emission offers  additional
means for neutron rejection. Secondary particles produced by neutron scattering are too massive to
reach the critical speed for Cherenkov emission; the same is true for scattered protons.
Neutron capture, instead, requires neutron
thermalisation; even if the neutron was detected through this process, its slow speed would
guarantee an effective TOF selection. 
Basically, the only possible source of background in presence
of a TOF-Cherenkov radiator comes from neutron-induced gamma rays that are not time-correlated 
or moderately correlated to the proton beam. \\
%
The locally constant background measured in the test with 63 MeV protons (figure \ref{fig:profiles}a)
provides an experimental evidence that the noise is not time-correlated in the region of interest for
PG monitoring. Its frequency ($f_{noise}$)
was estimated as the integral counts in the constant background divided by the acquisition duration,
resulting in $f_{noise}$~=~1.46 $\pm$ 0.02 Hz. Under the assumption that the  noise is mostly due to
the SiPM dark count rate ($DCR_{SiPM}$), $f_{noise}$ should be equal to the rate of random
coincidences $DCR_{coinc}$ between proton triggers in the diamond ($f_{dia}$) and  $DCR_{SiPM}$.
$DCR_{coinc}$ can be estimated as:
\begin{linenomath*} 
\begin{equation} 
DCR_{coinc}=f_{dia} \times DCR_{SiPM} \times 2\tau \label{eq:dcr} 
\end{equation}
\end{linenomath*} 
where $2\tau$ is the coincidence window width of 20 ns. From the proton beam intensity of 0.025
$\pm$ 0.002 p/bunch and the beam frequency of 25 MHz, $f_{dia}$~=~625 $\pm$ 50 kHz. The
intrinsic dark count of the SiPM, instead, was measured on the bench for the same threshold level of
 6 p.e., and then corrected for the temperature difference between the laboratory and the MEDICYC
experimental room; a value $DCR_{SiPM}$~=~302 $\pm$ 219 Hz was estimated. Thus, according to
equation \ref{eq:dcr},  $DCR_{coinc}$~=~3.8 $\pm$ 3.1 Hz. Despite the large experimental errors
involved in the evaluation  of $DCR_{coinc}$, its compatibility with $f_{noise}$~=~1.46 $\pm$ 0.02
Hz suggest that, the detection of particles other than PGs, if occurring, is certainly negligible in this
time window.\\ 
A further proof of this observation is given by data presented in figure \ref{fig:p1_thintarget},
left. In the experiment with 148 MeV protons, the PG module included a 6$\times$6 mm$^{2}$ SiPM that
ensured a better optical photon collection efficiency with respect to the 3$\times$3 mm$^{2}$ SiPM
used in the 63 MeV experiment. This allowed a higher threshold of 10 p.e. to be set, making the
level of SiPM dark count rate negligible with respect to the data acquisition rate. The histogram in
figure \ref{fig:p1_thintarget}, left was built with a 15 ns coincidence window, but no data was
acquired outside the 8 ns window defined by the bunch width. In summary, when the SiPM threshold is
set high enough to cut-off the dark count contribution, the Cherenkov-based gamma detector is
rather insensitive to neutron-associated background, and therefore 
the optimum candidate for the construction of a high sensitivity, fast PG detector. \\
Still, one last source of background remains in the collected data.
Figure \ref{fig:2spectra}, showed two bumps in the  TOF distributions, respectively in the
ranges ($-$1.0~$\div$~0.5) ns and (2.9~$\div$~4.1) ns for detectors A and B. The origin of these
events can be understood by TOF considerations: they cannot originate from in the target
since the upstream detector (det A) measures for them a TOF smaller than the one measured 
for PGs originating from the target. 
Since these particles cannot be faster than PGs, they must come from upstream x the target. 
The TOF of these background events is rather well defined (suggesting a localized spatial origin)
and it is compatible (for both the upstream and the downstream detectors) with  protons 
directly travelling from the diamond detector to the PG module.
Moreover, they appear in the same time-position of protons that are directly detected by the SiPM;
the latter have a very particular shape and they can be easily identified (and rejected) by pulse-shape
analysis. 
However, since protons of such energies cannot trigger Cherenkov production in the PbF$_2$, they must
be converted locally into PGs in order to maintain their TOF coherence. In summary, our hypothesis is 
that these contributions are due to protons scattering in the diamond PCB board and
interacting in the Cherenkov radiator holder made of Polyoxymethylene ($(CH_{2}O)_{n}$). The
chemical composition of Polyoxymethylene  is  very close to that of PMMA, and results in the local
production of PGs that are indistinguishable from the actual signal. Currently this effect cannot 
be rejected in the data, but it can be avoided in the future by improving the crystal holder design 
(or removing it altogether). We do not expect the PG conversion to take place in the crystal itself 
as  we have not observed this effect in the 63 MeV experiment, for which the crystal was simply wrapped
 in a black tape, with no holders. Still, this hypothesis requires further experimental verification.
In the light of this observation, it should also be observed that the scattered proton background must also
affect data in figure \ref{fig:p1_thintarget}.
\section*{Discussion} Prompt gamma emission is a rare physical phenomenon, therefore, increasing
detector sensitivity is  of utmost importance to achieve treatment monitoring in real time with
PG-based systems. Being able to measure the proton range \textit{in vivo} with a limited PG
statistics essentially means that the treatment can be verified very quickly at its very beginning
and avoid unwanted over-irradiation of the healthy tissue.\\
Our work focuses on two main aspects: proposing a new PG imaging technique (PGTI) to improve the
proton range measurement sensitivity, and conceiving an innovative PG imaging detector with high
detection efficiency and high time resolution. The potential performances of PGTI have been
discussed in a previous paper for different operating conditions. 
Here we focussed on the experimental feasibility of the PGTI technique in SPR in order to characterise the
inherent performances of the proposed detection module.  
The SPR was realised in manual delivery mode since the clinical settings do not currently allow for such low intensities as they have no clinical  applications yet. For the same reason, the dose delivery algorithm does not currently offer the possibility of switching from SPR to clinical intensity. While no technical barriers were identified in the delivery of the SPR intensity neither in terms of feasibility nor in terms of operation speed, it was observed that the SPR intensity is too low compared to the S2C2 ionisation chambers'  sensitivity. This would most probably require, in the future, the development of a  dose monitoring system (e.g. a diamond beam monitor) dedicated to SPR as well as a new delivery software in addition to the current IBA "blind golfer" algorithm.\\
For 63 MeV protons we achieved, in SPR, a proton range sensitivity of 4 mm (at 2$\sigma$) with an
unprecedentedly low statistics of only 600 PGs.  This value confirms  the 3 mm (at 2$\sigma$)
predicted by MC simulation\cite{jacquet_time--flight-based_2021} with 3000 PG events acquired. From these
simulations, it can also be estimated that 600 PGs would correspond to 2~$\times$~10$^{6}$ incident 
protons for the full 30-channels TIARA prototype (0.6\% detection efficiency).
This results paves the way to the use of the TIARA detector at the very beginning of the session 
to position the patient and/or verify the most critical spot(s) while operating  
a reduction of the beam intensity to $\sim$ 1 p/bunch. The duration of this monitoring procedure
varies according to the time characteristic of the accelerator employed. For an accelerator such as the MEDICYC cyclotron, 
delivering 10$^{7}$ protons in SPR would require about 0.63 seconds according to theoretical calculations, whereas this time would be longer (31.6 seconds) for the S2C2 synchrocyclotron. 
The SPR approach should therefore be considered part of the patient set-up procedure (of the order
of 15 minutes in the clinical practice) rather than as the treatment itself, for which PGTI should 
rather be implemented at nominal intensities (e.g. from $\sim$2000 to $\sim$2 million of p/bunch for the S2C2).\\
In the experiment with 148 MeV protons, we could compare the TOF distributions obtained with the
detector in two different positions. We qualitatively showed that using multiple detector configurations is
pivotal to obtain a uniform and increased detection efficiency (and eventually sensitivity)
throughout the proton range. This requires the use of a dedicated reconstruction algorithm as PGTI
in order to correct for the non-linearities introduced by the PG TOF term. This correction may not
seem necessary when using conventional gamma detectors but, when using a detection system with 235
ps (FWHM) time resolution, it is essential in order to fully exploit its potential precision.\\
PGTI therefore goes hand in hand with the development of new detectors with optimised time resolution and
detection efficiency. The most recent gamma detection module, composed of a 2~cm$^{3}$ PbF$_{2}$ and
a 6$\times$6 mm$^{2}$ MPPC, has shown a time resolution below 167 ps (FWHM) for 148 MeV protons
irradiations. With a larger photodetector surface compared to the previous module, it was possible
to set a higher detection threshold thus ensuring that no dark count events were acquired. A high
detection efficiency is guaranteed by the lack of a collimation system and by the optimisation of
the SNR as Cherenkov radiators are rather insensitive to background particles (mainly neutrons). 
We are currently working on the detector packaging optimisation to avoid the detection of scattered
protons and conceiving a mechanical system to hold multiple modules all around the patient. The latter
should be able to cope with the patient table, by either building a sort of helmet for the patient, or
placing some of the detectors behind the table. This second approach is possible as PGT/PGTI is not 
very sensitive to Compton scattering: scattered PGs are only very slightly delayed (few ps at worst) and
they maintain their temporal coherence (c.f. figure \ref{fig:neutrons}). \\ 
This work was carried out under the hypothesis that every single proton could be tagged
in time by selecting only 1-proton signals during the analysis. 
Experimentally, and with the current design of our beam monitor, this would require lowering 
the beam intensity to less than one proton per bunch\cite{dauvergne_role_2020} in order to minimise the
probability of 2- and 3-proton events; an approach that  would further increase the duration of
the monitoring procedure. 
In order to overcome this limitation, different solutions (software and hardware) are under investigation to tag in time each proton in the bunch. We are conceiving a dedicated algorithm that exploits the  increased rise time and the 
different shapes of 2-, 3-, 4-proton signals to extract separate time stamps. The precision of these time stamps would be worse 
than those obtained for 1-proton signals (397 ps FWHM at S2C2), but they would still be more precise than the 2.7 ns FWHM CTR
expected at nominal intensity for both PGT and PGTI.
At the same time, we are developing a large area, multi-channel diamond-based beam monitor\cite{gallin-martel_large_2018, dauvergne_role_2020} that would not only allow to overcome the size limitation of our current prototype, but also to 
tag in time multiple protons' signals with the same precision of single protons\cite{gallin-martel_large_2018,
sacchi_test_2020}. The combination of a multi-channel monitor and a dedicated time tagging algorithm could result in a further extension of the beam intensity for the "single proton" regime.  Still, once the protons' time stamps are available a dedicated algorithm should be developed to iteratively (or through ML) identify the very proton that has produced the detected PG. As a result of this procedure, some events will have a degraded  time-, and therefore space-resolution. A new assessment of the technique sensitivity will be therefore necessary in this scenario, but we expect to perform better, by design, than with  nominal intensities by either PGT or PGTI. \\
Finally, it should be kept in mind that the SPR is a possibility, not a requirement for PGTI and for our detector. 
Our approach would allow to perform an \textit{in-vivo} control of the patient set-up at the beginning of the treatment by verifying his/her anatomy, and then it could be used during the whole treatment at nominal intensities with performances 
that are not worse than PGT but with the advantage of employing Cherenkov detectors.
At clinical intensity (e.g. from $\sim$2000 p/bunch for the S2C2),  the loss
in time resolution could be compensated by the increased acquisition statistics (see Jacquet et
al.\cite{jacquet_time--flight-based_2021} for details) to achieve similar sensitivities in the
proton range measurement. 
At even higher intensities (the maximum intensity achievable with S2C2 is of $\sim$36$\times$10$^{6}$~p/bunch),
Cherenkov radiators offer very good perspectives to sustain
high count rates. The time-scale of Cherenkov process is of the order of the ps (to be compared to
tenths of ns at best for conventional scintillators), resulting in a negligible dead-time, with the
signal duration essentially given by the recharge time of the SiPM microcell. The latter can be
cut-off to few ns with the appropriate electronics. In fact, the low-light output of the Cherenkov
process ensures that only a few microcells per PG  are activated, with the others available for the
next event. It is therefore realistic to design a Cherenkov module that can sustain count rates up
to $\sim$100 MHz per channel. At these extreme count rates, however, the design of an electronic board capable
of tagging in time each PG is challenging and, at some point,  
different approches as the calculation of the center of gravity of the PG distribution
\cite{jacquet_time--flight-based_2021} should be used. For our final 30-channels prototype, we are
conceiving a dedicated electronic, which will be based on digital TDCs, in order to handle the different
 regimes.
%
\section*{Methods}
\subsection{Single proton regime}
For the 63 MeV experiment at the MEDICYC facility, the beam intensity was arbitrarily set to
obtain a negligible ratio of 2-protons signals at the diamond level. 
The MEDICYC cyclotron is already calibrated to work down to a nominal intensity of 0.1 p/bunch. \\
At the S2C2 synchrocyclotron, the beam intensity depends on two parameters: the voltage of the Dees (V$_{Dee}$) and  the S2C2 collimation slit opening.  V$_{Dee}$ is given as a percentage of the maximum value. In the clinical practice the system calibration is performed for  V$_{Dee}>$66.49\%. In this work, the  "effective" SPR at the beam monitor level required to set a V$_{Dee}$ of 65\%.
The slit opening, instead,  was set to the minimal  value of 1 mm. The spot integrity was verified in these conditions and no modifications were detected with respect to the clinical mode. \\
The SPR was  performed in "manual delivery mode" for feasibility and safety reasons, so as  to not corrupt the "clinical site configuration" of S2C2 which is extensively certified and validated for clinical purposes.
This configuration, in fact, does not enable the SPR, as these intensities have no clinical application at this time and any modification of the settings would require the complete recalibration of the facility and a double validation (from both IBA and the customer).
\subsection{Effective beam intensity}
The effective beam intensity at the beam monitor level was calculated \textit{a posteriori} taking
into account Poisson statistics. The diamond energy distribution was integrated in the regions
corresponding to 0 and 1 proton signals to obtain the probability of having zero (P(0)) or one (P(1))
protons in the bunch. The ratio P(1)/P(0) provides the $\lambda$ parameter of the Poisson distribution
describing proton delivery, which corresponds to the average number of protons per bunch. \\
Nevertheless, the intensity values calculated in this work do not  correspond to the actual intensity set
at the accelerator level. For the MEDICYC cyclotron, the calculation is biased by the presence of a
collimator, whereas, for the S2C2 synchrocyclotron, the beam monitor covered only $\sim$20\% of the beam surface, meaning 
that the actual beam intensity was of the order of 4.7 p/bunch .
\subsection{Intrinsic detection efficiency}
A  Geant4\cite{allison_recent_2016} simulation (version 10.4.p02)   of the optical properties of the TIARA  module,
based on the UNIFIED model and the QGSP-BIC-EMY
physics list\cite{wronska_prompt-gamma_2021},  was performed to establish the module intrinsic  detection efficiency as a function
of the incident PG energy. The efficiency was  computed  as the fraction of PGs depositing more than 100 keV in
the crystal and resulting in more than N$_{th}$ p.e. reaching the SiPM (with N$_{th}$ being the threshold expressed in p.e). The  SiPM photodetection efficiency was also taken into account. 
Simulated data were then fitted with an analytical function (given by the sum of a sigmoid and a first degree polynomial) that was exploited to take into account the detector response in other MC simulations carried out for this work.
The functions obtained for different values of N$_{th}$ are shown in figure \ref{fig:det_eff}.
\subsection{Generation of reference profiles} 
Reference profiles are built from MC simulations of
the reference geometry. The Geant4.10.4.p02 version with the QGSP-BIC-EMY
physics list was used to generate the PG time stamps on a detection
surface of the same size as the gamma-detector module in order to take into account its geometrical
detection efficiency. \\
The detector response was subsequently taken into account by considering the detection probability of each simulated event as defined by the analytical function presented in figure \ref{fig:det_eff} for a threshold of 6 p.e (orange curve).
The system TOF resolution was then included by
convolving the data with a gaussian distribution of 315 ps FWHM (i.e. the experimentally
measured value).\\ 
This multi-step approach was chosen to reduce the computing time, as PG
generation is a  rare phenomenon and the simulation of optical photon propagation and interactions
in Geant4 are time-consuming. The excellent agreement between the experimental and simulated
reference profiles in figure \ref{fig:profiles} (b and d) is an indirect validation of the
simulation procedure.
\subsection{Measurement of the distance between two PG profiles} After background rejection, the
procedure used to measure the distance between two PG profiles is the same for PGT and PGTI
(reconstructed) profiles. 
First, the reference X value $x_{ref}$ (either in the unit of time or
space for PGT or PGTI respectively) is defined as the distal maximum in the simulated reference
profile. Then, each experimental PG profile and the simulated PG profile are integrated to exploit
the noise-filtering properties of the integral function. The difference $d_{i}$ between each
experimental integrated PG profile ($f_{i}(x)$) and the integrated reference profile ($f_{ref}(x)$)
is calculated as $d_{i}=f^{-1}(y_{ref})-x_{ref}$, where $y_{ref}=f_{ref}(x_{ref}$). This method is
described in detail in Marcatili et al. \cite{marcatili_ultra-fast_2020}. 
\subsection{Errors for the sensitivity plot} The error on the profile fall-off position (cf. figure
\ref{fig:sensitivity}) is determined from toy experiments, using the bootstrap technique. For each
experimental TOF profile, 5000 sub-samples (toy experiments) including from 30 to 135 PGs (in steps
of 15 PGs) were extracted, for a total of 8 sets of 5000 data samples per profile. The size of
sub-samples was kept small to limit their statistical dependency. The 5000 sub-samples were then
used to estimate the 1$\sigma$ and 2$\sigma$ statistical errors on the difference $d_{i}$ between
the toy experiment profile and the reference profile, and to obtain their dependency on the number
$N$ of PG events in the profile. This dependencies varies as $k/\sqrt{N}$ (where $k$ is a constant)
and allows to extrapolate the experimental statistical errors at 1$\sigma$ and 2$\sigma$ for the PG
statistics available in the current experiment.
\subsection{Background subtraction in PGTI distributions} The background in PGT distributions is
flat as it is generated from events that are not time-correlated to the PG signal. When the PGT
distributions are reconstructed, the flat background is non-linearly transformed acquiring a
complex, non constant shape. A model of the PGTI background is built by reconstructing a constant
signal according to equation \ref{eq:pgti}. The model is then fitted on the reconstructed TOF
histogram and subtracted the background.\\ It should be noted, that a more straightforward method
would have been to subtract the flat background from the PGT profile before reconstruction. However,
with the aim of implementing here an event-by-event reconstruction that could be performed as data
are acquired, we performed the analysis under the assumption that the background level is not known
at reconstruction. 
\section*{Data availability} The datasets used and/or analysed during the current study are
available from the corresponding author on reasonable request.
%

\section*{Acknowledgements} The authors would like to thank Sebastien Henrotin  and Cedric
Osterrieth for kindly assisting with the tuning of the single proton regime at the ProteusOne
S2C2.\\ This work was partially supported by the ANR (project ANR-15-IDEX-02), INSERM Cancer (TIARA
project), the LABEX PRIMES (ANR-11-LABX-0063) of Universit\'{e} de Lyon and by the European Union
(ERC  project PGTI,  grant number 101040381). Views and opinions expressed are however those of the authors only and do not
necessarily reflect those of the European Union or the European Research Council Executive Agency.
Neither the European Union nor the granting authority can be held responsible for them.

\section*{Author contributions statement} M.J. and S.A. analysed the results; S.M., M.L.G.M. and
Y.B. supervised the analysis; S.M., M.J., D.M., J.H., J.P.H., C.M and M.D. conceived the different
experiments; M.J., M.L.G.M., S.A., A.A. and D.M. conducted the experiments; J.E. and FS calibrated
the S2C2 accelerator for the single proton regime. C.H. and L.G.M. developed the electronics; J.F.M.
conceived the detector integration. S.M. wrote the manuscript. All authors reviewed the manuscript.
\end{document}